%% file: DESY-00-133.tex
\documentclass[a4paper,12pt,twoside]{article}
\usepackage[english]{babel}
\usepackage{graphics}
\usepackage{epsfig}
\usepackage{verbatim}
\usepackage{rotating}
\usepackage[centertags]{amsmath}
\usepackage{array}
\usepackage{multirow}
\usepackage{supertabular}
\usepackage{axodraw}
\usepackage{dcolumn}
\usepackage{./styles/mcite}
\usepackage{xspace}
\usepackage{graphicx}
\usepackage{amsfonts}
\usepackage{amssymb}
\newcommand{\ptmiss}{{\mbox{$\not\hspace{-.55ex}{P}_T$}}}
\newcommand{\rpvio}{{\not\hspace{-.55ex}{R}_p}}
\input zeuspap_def.tex

\input zeuspap_fmt.tex

\begin{document}

\title {\bf\LARGE A search for resonance decays to $\nubar$-jet  
in $e^+p$ scattering at HERA}

\author{ZEUS Collaboration}

\date{}

\maketitle

\begin{abstract}
A study of the $\bar\nu$-jet mass spectrum in
$e^{+}p\rightarrow\bar\nu X$ events at center-of-mass energy 300 GeV
has been performed with the ZEUS detector at HERA using
an integrated luminosity of $47.7\mbox{pb}^{-1}$.
The mass spectrum is in good agreement with that 
expected from Standard Model processes
over the $\bar\nu$-jet mass range studied.
No significant excess
attributable to the decay of a narrow resonance is observed.
By using both $e^{+}p\rightarrow e^{+} X$ and 
$e^{+}p\rightarrow\bar\nu X$ data,
mass-dependent limits are set on the $s$-channel production of 
scalar and vector resonant states.
Couplings to first-generation quarks are considered
and limits are presented as a function of the $e^+q$ and $\nubar q$
branching ratios. 
These limits are used to constrain the production of leptoquarks and
$R$-parity violating squarks.

\end{abstract}

\vspace{-14.5cm}
\begin{flushleft}
\tt DESY 00-133 \\
September 2000 \\
\end{flushleft}

{\hspace*{-0mm} } \pagestyle{plain} \thispagestyle{empty} \newpage 

\topmargin-1.cm                                                                                    
\evensidemargin-0.3cm                                                                              
\oddsidemargin-0.3cm                                                                               
\textwidth 16.cm                                                                                   
\textheight 680pt                                                                                  
\parindent0.cm                                                                                     
\parskip0.3cm plus0.05cm minus0.05cm                                                               
\def\3{\ss}                                                                                        
\newcommand{\address}{ }                                                                           
\pagenumbering{Roman}                                                                              

\begin{center}                                                                                     
{                      \Large  The ZEUS Collaboration              }                               
\end{center}                                                                                       
  J.~Breitweg,                                                                                     
  S.~Chekanov,                                                                                     
  M.~Derrick,                                                                                      
  D.~Krakauer,                                                                                     
  S.~Magill,                                                                                       
  B.~Musgrave,                                                                                     
  A.~Pellegrino,                                                                                   
  J.~Repond,                                                                                       
  R.~Stanek,                                                                                       
  R.~Yoshida\\                                                                                     
 {\it Argonne National Laboratory, Argonne, IL, USA}~$^{p}$
% ...
\newpage                                                                                           
% ...
                                                           %
                                                           %
\newpage   
                                                           %
                                                           %
\begin{tabular}[h]{rp{14cm}}                                                                       
$^{a}$ &  supported by the Natural Sciences and Engineering Research                               
          Council of Canada (NSERC)  \\                                                            
$^{b}$ &  supported by the FCAR of Qu\'ebec, Canada  \\                                            
$^{c}$ &  supported by the German Federal Ministry for Education and                               
          Science, Research and Technology (BMBF), under contract                                  
          numbers 057BN19P, 057FR19P, 057HH19P, 057HH29P, 057SI75I \\                              
$^{d}$ &  supported by the MINERVA Gesellschaft f\"ur Forschung GmbH, the                          
          German Israeli Foundation, the Israel Science Foundation, the                            
          U.S.-Israel Binational Science Foundation, the Israel                                    
          Ministry of Science and the Benozyio Center for High Energy                              
          Physics\\                                                                                
$^{e}$ &  supported by the German-Israeli Foundation, the Israel Science                           
          Foundation, the U.S.-Israel Binational Science Foundation, and by                        
          the Israel Ministry of Science \\                                                        
$^{f}$ &  supported by the Italian National Institute for Nuclear Physics                          
          (INFN) \\                                                                                
$^{g}$ &  supported by the Japanese Ministry of Education, Science and                             
          Culture (the Monbusho) and its grants for Scientific Research \\                         
$^{h}$ &  supported by the Korean Ministry of Education and Korea Science                          
          and Engineering Foundation  \\                                                           
$^{i}$ &  supported by the Netherlands Foundation for Research on                                  
          Matter (FOM) \\                                                                          
$^{j}$ &  supported by the Polish State Committee for Scientific Research,                         
          grant No. 112/E-356/SPUB/DESY/P03/DZ 3/99, 620/E-77/SPUB/DESY/P-03/                      
          DZ 1/99, 2P03B03216, 2P03B04616, 2P03B03517, and by the German                           
          Federal Ministry of Education and Science, Research and Technology (BMBF)\\              
$^{k}$ &  supported by the Polish State Committee for Scientific                                   
          Research (grant No. 2P03B08614 and 2P03B06116) \\                                        
$^{l}$ &  partially supported by the German Federal Ministry for                                   
          Education and Science, Research and Technology (BMBF)  \\                                
$^{m}$ &  supported by the Fund for Fundamental Research of Russian Ministry                       
          for Science and Edu\-cation and by the German Federal Ministry for                       
          Education and Science, Research and Technology (BMBF) \\                                 
$^{n}$ &  supported by the Spanish Ministry of Education                                           
          and Science through funds provided by CICYT \\                                           
$^{o}$ &  supported by the Particle Physics and                                                    
          Astronomy Research Council \\                                                            
$^{p}$ &  supported by the US Department of Energy \\                                              
$^{q}$ &  supported by the US National Science Foundation                                          
\end{tabular}                                                                                      
                                                           %
\newpage

\newpage

\topmargin-1.5cm
\evensidemargin-0.3cm
\oddsidemargin-0.3cm
\textwidth 16.cm
\textheight 650pt
\parindent0.cm
\parskip0.3cm plus0.05cm minus0.05cm

\pagenumbering{arabic} 
\setcounter{page}{1}

\section{Introduction}

\label{sec-int}

A number of extensions of the Standard Model of elementary particles
predict the existence of electron-quark resonant states at high mass.
Such states include
leptoquarks (LQs)~\cite{buch} and $R$-parity violating ($\rpvio$)
squarks~\cite{squarks}.
The corresponding production processes could give 
a large cross section for high-mass
$\nubar$-jet or $e^+$-jet events. 

This paper presents an analysis of the ZEUS data aimed at searching for
high-mass scalar and vector resonant states decaying into an antineutrino
plus a jet. 
A similar search in the $e^+$-jet final states with the ZEUS
data was published previously~\cite{ncref}.
To avoid the constraints from 
a specific model, minimal assumptions are made about the properties of 
the resonant state. 

This analysis uses events whose observed final state has large missing 
transverse momentum and at least one jet. 
These event characteristics 
correspond to an outgoing antineutrino and a scattered quark in
$e^{+}p\rightarrow\nubar X$ scattering. The data-selection and event-reconstruction techniques are similar to those used for measuring the
charged current (CC) cross section~\cite{ccpaper}. 
The $\nubar$-jet invariant mass is
calculated from the energies and angles of the final-state antineutrino and 
jet:

\begin{equation}
M_{\nu j}^{2}=2E_{\nu}E_{jet}(1-\cos\xi) \label{invm_eqn}
\end{equation}

\noindent where $E_{\nu}$ and $E_{jet}$ are the 
energies of the scattered
antineutrino and jet (assumed massless), respectively.  
The angle $\xi$\ is the laboratory-frame
opening angle between the jet and the antineutrino. 
Since the antineutrino escapes detection, its momentum is
deduced from all observed final-state particles by assuming conservation of
energy-momentum in the event. 

In the following sections, the expectations of antineutrino-jet
final states from the Standard Model (SM) 
and from models that predict resonant states are first reviewed.
After a summary of experimental conditions and data selection, 
the analysis is described and the reconstructed mass spectrum is presented.
Since there is no evidence for a narrow resonance in either
the $\nubar$-jet or the previously published 
$e^+$-jet mass spectra, limits are set on the
production of positron-quark resonant states using both data sets.
The application of these limits to LQ and squark production is then discussed.

\section{Signal and background expectations}

\begin{figure}[ptb]
\centerline{\epsfxsize=8cm  \epsfysize=10cm
\epsfbox{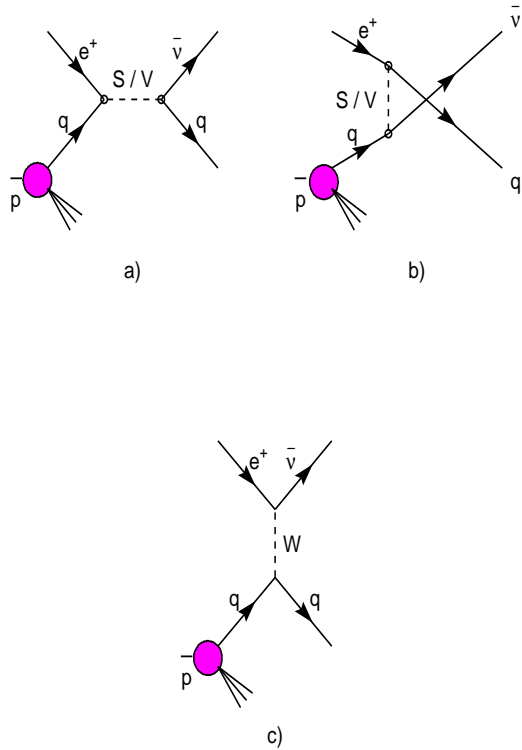}}
\caption{
Processes with $\nubar$-jet final states in $e^{+}p$ collisions.
A scalar (S) or vector (V) intermediate state can be formed
via a) $s$-channel or b) $u$-channel exchange.
Weak charged current scattering c)
forms the primary background to these processes.
}
\label{fig_fyn}
\end{figure}

High-mass $\nubar$-jet final states can be formed either through SM
mechanisms or via processes that produce lepton-quark 
resonances.
Figure~\ref{fig_fyn} shows scattering mechanisms producing
such final states in $e^+p$ collisions.
The CC scattering mechanism shown in Fig.~\ref{fig_fyn}c
forms the primary background in this search.
Neutral current (NC) and photoproduction processes
form negligible backgrounds since neither produces events with a large
observed final-state momentum imbalance.

\subsection{Standard Model expectations}

The kinematic variables used to describe the process 
$e^+ p \rightarrow \nubar X$ are:
\begin{align}
Q^{2} &  =-q^{2}=-(k-k^{\prime})^{2}\\
x  &  =\frac{Q^{2}}{2P\cdot q}\\
y  &  =\frac{q\cdot P}{k\cdot P}
\end{align}

\noindent where $P$ is the four-momentum of the incoming proton, and $k$ and
$k^{\prime}$ are the four-momenta of the incoming positron and the outgoing
antineutrino, respectively. 
These variables are related by $Q^{2}=sxy$. The quantity
$x$ is interpreted as the fraction of the proton momentum carried
by the struck quark, and $y$ measures the fractional energy transferred 
by the $W$ in the CC process. 

Assuming no QED or QCD radiation, the mass of the $\nubar q$ system is
related to $x$ via
\begin{equation}
\label{eq:mass}
M^2 = sx
\end{equation}
and the scattering angle, $\theta^{\ast}$, of the outgoing antineutrino
relative to the beam positron, as viewed in the $\nubar q$ center-of-mass
system, is related to $y$ via 
\begin{equation}
\label{eq:cost}
\cos \theta^{\ast}=1-2y \; .
\end{equation}

In leading-order electroweak theory, the CC cross section can be expressed as:

\begin{equation}
\frac{d^{2} \sigma^{CC}(e^{+}p)}{dxdy} = \frac{s G_{F}^{2}}{4 \pi}\left(  \frac
{M_{W}^{2}}{M_{W}^{2} + xys}\right)  ^{2}\left[  Y_{+}F^{CC}_{2} -
Y_{-}xF^{CC}_{3} - y^{2}F^{CC}_{L}\right]
\label{eqn_crossec}
\end{equation}

\noindent where $G_{F}$ is the Fermi constant, $M_{W}$ is the mass of the $W$
boson, and $Y_{\pm}=1\pm(1-y)^{2}$. \ The proton structure functions
$F_{2}^{CC}$ and $xF_{3}^{CC}$, in leading-order (LO) QCD, measure respectively sums and differences of quark and antiquark parton momentum 
densities~\cite{strfunc}.
The longitudinal structure function, $F_{L}^{CC}$, contributes negligibly to
this cross section except at $y$ near 
1~\cite{ccpaper}. \ In the region of high mass
($x\rightarrow1$) the structure functions $F_{2}^{CC}$ and $xF_{3}^{CC}$ are
dominated by the valence quark distributions in the proton. \ For $e^{+}p$
collisions, the scattering from down quarks
dominates the cross section. 
The CC cross section peaks at small $y$, which 
leads to a $\cos\theta^{\ast}$ distribution rising
toward $\cos\theta^{\ast}=1$.

The largest uncertainty in the 
CC cross-section prediction arises from 
the parton densities of the proton. The parton density functions (PDF) are
parameterizations which, at high $x$, are determined primarily 
from measurements made in fixed-target deep inelastic scattering (DIS) 
experiments. In the high-mass range ($x\approx0.6$ corresponding to a 
$\nubar$-jet mass of $230\,\mbox{GeV}$), the PDFs introduce an uncertainty 
of $\approx25\%$ in the predicted $e^+p$ CC cross section~\cite{botje}. 
It should be noted that recent studies of PDFs suggest that the
$d$-quark density in the proton has been systematically underestimated for
$x>0.3$~\cite{botje,melnit,bodek,cteq5}.  
As an example, Yang and Bodek \cite{bodek} propose a correction to 
the $d/u$\ quark density ratio in the MRS(R2) PDF~\cite{MRSR2}:
\begin{equation}
\delta\left(  \frac{d}{u}\right)  =0.1x(x+1)\label{eqn_by}
\end{equation}
\noindent which fits the available data better. \ When
this correction is applied to the CTEQ4D PDFs~\cite{cteq4}, the increase in the
predicted CC cross section (and the corresponding number of high-mass $\nubar
$-jet events) ranges from $1.0\%$ at $x=0.1$ to  $60\%$ at $x=0.6$.
More recent PDF parametrizations~\cite{botje,cteq5,MRST}, 
agree well with the corrected CTEQ4 for $x$ up to 0.7.

\subsection{High-mass resonant states}

If a high-mass resonant state were produced at HERA, it could have
a final-state signature similar to NC or CC DIS.
Electron-quark states which couple to a single
quark generation and preserve lepton flavor are considered here. 
For $e^+p$ scattering, first-generation couplings of the form
$e^+u$,~$e^+ d$,~$e^+ \bar u$ and $e^+ \bar d$ can be defined.

These states are classified using the fermion number $F=L+3B$,
where $L$ is the lepton number and $B$ is the baryon number of the state.
The coupling of positrons to quarks ($e^+u $ and $e^+ d $) requires $F=0$ and the coupling of positrons to antiquarks ($e^+ \bar u $ and $e^+ \bar d $) requires $F=-2$.
In $e^+p$ scattering, the $F=0$ 
states couple to the valence quarks of the proton and, for the same coupling,
would have a significantly
larger cross section than would the $F=-2$ states.

\begin{table}[ptb]
\begin{center}
\begin{tabular}
[c]{|c|c|c||c|c|c|}\hline
\multicolumn{3}{|c||}{Scalar} & \multicolumn{3}{|c|}{Vector} \\ \hline
Resonance   & Charge   & Decay   & Resonance & Charge & Decay \\ \hline
$S_{e^+ u}$ & 5/3 & $e^+ u$ & $V_{e^+ u}$ & 5/3 & $e^+ u$ \\
$S_{e^+ d}$ & 2/3 & $e^+ d$ & $V_{e^+ d}$ & 2/3 & $e^+ d$ \\ 
 & & $\bar \nu u$   & & & $\bar \nu u$ \\ \hline \hline
$S_{e^+ \bar u}$ & 1/3 & $e^+ \bar u$ & $V_{e^+ \bar u}$ & 1/3 & $e^+ \bar u$ \\
& & $\bar \nu \bar d$ &  &  &$\bar \nu \bar d$ \\
$S_{e^+ \bar  d}$ & 4/3 & $e^+ \bar d$ & $V_{e^+ \bar d}$ & 4/3 & $e^+ \bar d$ \\ \hline
\end{tabular}
\end{center}
\caption{Possible first-generation scalar and vector
resonant states in $e^+p$ scattering.
The top half of the table lists color-triplet states with fermion number
$F=L+3B=0$, while the bottom half lists those with $F=-2$. The left and
right sets of columns list scalars and vectors, respectively. The $e^+ d$ and 
$e^+ \bar u$ states can decay to both $\nubar q$ and $e^+q$.
For the other states, only $e^+q$ decays are allowed since a $\nubar q$ 
decay would violate charge conservation.
}
\label{tab_resst}
\end{table}

Table~\ref{tab_resst} lists the 8 scalar and vector resonant states 
considered here, along with their 
charges and relevant decay modes. The $e^+ \bar u$ and 
$e^+ d$ states would produce both $e^+ q$ and $\bar \nu q$ final 
states, which correspond to NC and CC event topologies, respectively. 
The other states would decay only to  
$e^+ q$ since a $\nubar q$ mode would violate charge conservation. 
Some physics models incorporating high-mass  resonances 
predict additional decay channels with final-state topologies different
from DIS events. 
The branching ratios of each resonance into $e^+ q$, $\bar \nu q$ and other
final states are treated as free parameters except when specific models
with restricted branching ratios are considered.

In general, high-mass states formed by $e^+p$ collisions can have 
a combination of left- 
($\lambda _L$) and right- ($\lambda _R$) handed couplings. Because decays to
right-handed antineutrinos must occur through left handed couplings, only 
left-handed coupled states ($\lambda _R=0$) are considered 
for $\nubar q$ decays.

If  a state with mass
$M_{e^+q}<\sqrt{s}$ exists, the $s$-channel mechanism (Fig.~1a)
would produce a resonance 
at $M_{\nu j}=M_{e^+q}$ in $\nubar q$ decays. 
Additional contributions to the $e^+p$ cross section come from
$u$-channel exchange (Fig.~1b) and the interference with $W$ 
exchange (Fig.~1c).
The total $e^+p \rightarrow \bar \nu X$ cross section with a resonance
contribution can be written as \cite{buch}

\begin{equation}
\frac{d^{2}\sigma(e^{+}p)}{dxdy}= \frac{d^{2}\sigma^{CC}}{dxdy} +
\frac{d^{2}\sigma^{Int}_{u/CC}}{dxdy} + 
\frac{d^{2}\sigma^{Int}_{s/CC}}{dxdy} + 
\frac{d^{2}\sigma_{u}}{dxdy} +
\frac{d^{2}\sigma_{s}}{dxdy}.
\label{eqn_lqcc}
\end{equation}

The first term on the right-hand side of Eq.~(\ref{eqn_lqcc}) represents 
the 
charged current contribution from the SM. The second (third)  term is the 
interference between the SM and $u$-channel ($s$-channel) exchange, 
and the 
fourth (fifth) term represents the $u$-channel ($s$-channel) exchange alone.
The contribution of a single vector or scalar state has two free
parameters: $M_{e^+q}$, the mass of the state and  $\lambda$,
its coupling to $e^{+}$-quark.
The $\cos\theta^{\ast}$ dependence of the state varies strongly for the 
different terms:
it is uniform for a scalar state produced in the $s$-channel or a vector 
state produced
in the $u$-channel, while it varies as $(1+\cos\theta^{\ast})^2$ for a vector 
state produced in the $s$-channel or a scalar state produced in 
the $u$-channel~\cite{buch}.

For the small couplings considered here, and if $M_{e^+q}<\sqrt{s}$,
the narrow resonance produced by the $s$-channel exchange would 
provide the dominant additional contribution over the SM background. 
The width of the $s$-channel resonance is given, e.g., for the
$S_{e^+q}$ with 50~\% branching to $\nubar q$, by

\begin{equation}
\Gamma _{e^+q} = \frac{M_{e^+q}}{16 \pi}(2\lambda ^2)
\end{equation}

so that if $\lambda ^2$ is sufficiently small, the production cross 
section 
can be approximated by integrating over the $s$-channel
contribution to the cross section. This leads to
the narrow-width approximation for the total cross section of 
a single state \cite{buch}:

\begin{equation}
\sigma^{NWA}=(J+1)\frac{\pi}{4s}\lambda^{2}q(x_{0}, M_{e^+q}^2) ,
\label{eqn_nwidth}
\end{equation}

where $q(x_{0}, M_{e^+q}^2)$ is the initial-state quark (or antiquark) 
momentum density in the proton
evaluated at $x_{0}=M_{e^+q}^{2}/s$ and at a virtuality scale of $M_{e^+q}^2$,
and $J$ is the spin of the state.
In the limit-setting procedure (Sect. \ref{sec_limits}), 
this cross section was
corrected for expected QED and QCD radiative effects. 
The effect of QED radiation on the resonant-state cross section was 
calculated and was found to decrease the cross section by $5-25\%$  as
$M_{e^+q}$ increases from $100 \rightarrow 290 \; \mbox{GeV}$.
For scalar resonant states, the QCD corrections~\cite{qcdcorr} 
raise the cross section by $20-30\%$ for $F=0$ resonances.
For $F=2$ states, the QCD corrections lower the cross section
by $5-30\%$ in the 200-290 GeV mass range.
No QCD corrections were applied to vector states because the calculation for 
such states is not renormalizable~\cite{Bluemlein}.

\begin{table}[ptb]
\begin{center}
\begin{tabular}
[c]{|c|c|c|c|c|c|}\hline
LQ species & Charge & F & Production & Decay & Branching ratio\\\hline
$V^{L}_{0}$ & -2/3 & 0 & $e_{L}{\bar d}_{R}$ & $e{\bar d}$ & 1/2 \\
&  &  & & $\nu{\bar u}$ & 1/2 \\ \hline
$S^{L}_{0}$ & -1/3 & 2 & $e_{L} u_{L}$ & $eu$ & 1/2\\
&  &  & & $\nu d$ & 1/2\\ \hline
\end{tabular}
\end{center}
\caption{First-generation leptoquark species considered in this analysis.
 The superscript $L$ denotes
chirality, while the subscript 0 indicates the weak isospin. The
electric charge, the production channel, and the allowed decay channels
are also displayed.  For positron beams, the charge changes sign, the 
helicity of the lepton is reversed, and the quarks and anti-quarks 
are interchanged. }
\label{tab_lqs}
\end{table}

\section{Resonant-state models}

In the absence of a clear resonance signal, 
limits can be placed on the production of states in models 
which predict a high-mass positron-quark resonance decaying to  
$e^+q$ or $\nubar q$. Two such models are considered: (1) leptoquark
(LQ) states with
$SU(3) \times SU(2) \times U(1)$ invariant couplings and 
(2) squark states found in $R$-parity violating supersymmetry
(SUSY) models.

\subsection{Leptoquarks}

For $SU(3)\times SU(2)\times U(1)$ invariant LQ couplings,
there are 14 possible LQ species \cite{buch}. 
Such leptoquarks
have no decay channels other than $e^+q$ or $\nubar q$.
Table~\ref{tab_lqs} lists those which have
equal branching ratios into $e^+q$ and $\nubar q$ decays.
These scalar and vector LQ species correspond to the $S_{e^+\bar u}$ and $V_{e^+d}$ resonant states, respectively, 
with branching ratios fixed to 
$\beta_{e^+ q} = \beta_{\nubar q} = 1/2$.

\subsection{SUSY}

\begin{table}[ptb]
\begin{center}
\begin{tabular}
[c]{|c|c|c|}\hline
Production & Decay & Resonance \\ \hline
 & $e^+ d$ & \\
$e^{+}+{d}\rightarrow\tilde{{u}}_j$ & $\chi ^0_i u_j$ & $S_{e^+d}$\\
 & $\chi ^+_i d_j$ & \\ \hline
&  $e^+\bar u$ & \\ 
$e^{+}+{\bar{u}}\rightarrow\tilde{{\bar{d}}}_k$ & $\nubar \bar d$ & $S_{e^+\bar u}$ \\
 & $\chi ^0_i \bar d_k$  & \\ \hline
\end{tabular}
\end{center}
\caption{Squarks predicted by SUSY that have $\rpvio$ decays into 
$e^+$-jet or $\nubar$-jet final states. Listed are the squark production 
mechanism and decay channel. The $k$ and $j$ subscripts indicate the 
squark generation.
Also shown is the corresponding resonant state from Table~\ref{tab_resst}. 
The decay modes with a $\chi^{+,0}_{i}$ are the $R$-parity--conserving decay 
modes which produce  neutralinos ($\chi ^0_i$) and charginos ($\chi ^+_i$). 
These undergo further decays into SM particles.
}
\label{tab_SUSY}
\end{table}

\begin{figure}[ptb]
\centerline{\epsfxsize=9cm  \epsfysize=12cm
\epsfbox{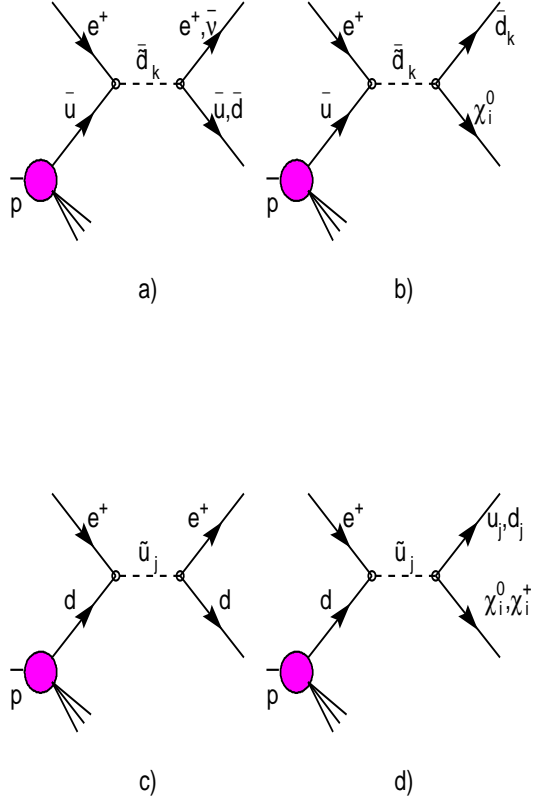}}\caption{
Lowest-order $s$-channel diagrams for first-generation squark production 
in $e^+p$ collisions at HERA. Diagrams a) and c) are the 
$\rpvio$ decays for $\tilde{{\bar{d}}}_{k}$ and  
$\tilde{u}_j$ squarks, respectively. The $R_p$-conserving decays
are shown in b) and d).
The decays of the charginos and neutralinos, 
$\chi ^0_i$ and $\chi ^+_i$, into SM particles depend on the parameters 
of the SUSY model and are not shown.
}
\label{fig_fyn_susy}
\end{figure}

In SUSY, conservation of baryon and lepton number is expressed in 
terms of $R$-parity, $R_p$. It is defined as $R_p=(-1)^{3B+L+2S}$, 
where $B$ is the baryon number, $L$ is the lepton number and $S$ is the 
spin of the particle.
Ordinary SM particles have $R_p=+1$ while their hypothetical supersymmetric
partners  have $R_p=-1$.
In versions of the theory in which $R$-parity is not conserved, 
squarks (the SUSY counterparts to quarks) 
have the same production mechanism as a generic scalar resonance.
The squark flavors listed in Table~\ref{tab_SUSY} have $\rpvio$ 
decays into lepton-jet final states. 
Figures~\ref{fig_fyn_susy}a and c show the $s$-channel diagrams for these 
squark decays.
The $\tilde{u}_j$ and the $\tilde{{\bar{d}}}_{k}$ squarks behave like
$S_{e^+d}$ and $S_{e^+\bar u}$ resonant states, respectively 
(see Table~\ref{tab_SUSY}), and the subscripts $j$ and $k$ denote the
squark generation. Three generations are possible, but it is assumed that only
a single generation has non-negligible coupling.
These squarks would also be expected to have $R_{p}$-conserving decays
into neutralinos ($\chi ^0_i$) and charginos ($\chi ^+_i$) 
(Figs.~\ref{fig_fyn_susy}b and d) with multi-jet
signatures different from $e^+$-jet and $\nubar$-jet.
A detailed discussion of these states, whose properties depend on 
many SUSY parameters, is beyond the scope of this paper.
The branching ratios of squarks into $e^+$-jet and $\nubar$-jet, as well as other final states, are therefore treated as free parameters in this paper.

\section{Experimental conditions}

During 1994-97, HERA collided protons of energy $E_{p}=820\,\mbox{GeV}$ with
positrons of energy $E_{e}=27.5\,\mbox{GeV}$. The integrated luminosity of the
data is $47.7\mbox{pb}^{-1}$. A detailed description of the
ZEUS detector can be found elsewhere~\cite{zeusdesc}. \ The primary components used
in the present analysis are the central tracking detector (CTD) positioned
in a 1.43 T solenoidal magnetic field, the
uranium-scintillator sampling calorimeter (CAL) and the luminosity 
detector (LUMI).

The CTD \cite{ctddesc} was used to establish an interaction vertex with 
a typical resolution of 3~cm in the beam direction for events considered in
this analysis. 
Energy deposits in the CAL \cite{caldesc} were used to measure the 
positron energy and hadronic energy. The CAL has three sections: the 
forward\footnote{The ZEUS coordinate system is right-handed with the $Z$ 
axis pointing in the direction of the proton beam (forward) and the $X$ axis 
pointing horizontally toward the center of HERA. The polar angle $\theta$
is defined with respect to the $Z$ axis.},
barrel, and rear calorimeters (FCAL, BCAL, and RCAL).
The FCAL and BCAL are segmented longitudinally 
into an electromagnetic section (EMC) and two hadronic sections  (HAC1, 2).  
The RCAL has one EMC and one HAC section.  The cell structure is formed by  
scintillator tiles.
The cells are arranged into towers consisting of $4$ EMC cells, a HAC1
cell and a HAC2 cell (in FCAL and BCAL).  The transverse dimensions
of the towers in FCAL 
are $20\times 20$~cm$^2$.  One tower is absent at the center of the FCAL
and RCAL to allow space for passage of the beams. 
Cells provide timing measurements with resolution better than 
1 ns for energy deposits above 4.5 GeV. Signal times are useful
for rejecting background from non-$ep$ sources and for determining the position
of the interaction vertex if tracking information is unavailable.

Under test beam conditions, the CAL has a resolution of  
$0.18/\sqrt{E(\mbox{GeV})}$ for positrons
hitting the center of a calorimeter cell, and 
$0.35/\sqrt{E(\mbox{GeV})}$ for single
hadrons. The events of interest in this analysis have only hadronic jets, 
which impact primarily in the FCAL. In simulations, 
the jet energy resolution for 
the FCAL is found to average 
$\sigma/E = 0.55/\sqrt{E(\mbox{GeV})}\oplus 0.02$~\cite{ncref}.

To reconstruct the hadronic system, corrections were applied
for inactive material in front of the calorimeter. 
The overall hadronic energy scales of the FCAL and BCAL are determined to 
within $2\%$ by examining the $P_T$ balance of NC DIS events~\cite{nchighq}.

The luminosity was measured from the rate of the bremsstrahlung process
$e^+p \rightarrow e^+p\gamma$ \cite{lumidesc}, and has an uncertainty of 
$1.6\%$.

A three-level trigger similar to the one used in the charged current
analysis was used to select events online~\cite{ccpaper}. 

\section{Event simulation}

Standard Model CC events were simulated using the 
HERACLES 4.6.2~\cite{heracles} program 
with the DJANGO 6 version 2.4~\cite{django} 
interface to the hadronization programs. 
First- and second-generation quarks are simulated, while third-generation 
quarks were ignored~\cite{ukatz} because of the large mass of
the top quark and the small off-diagonal elements of the 
CKM matrix. The
hadronic final state was simulated using the MEPS model in LEPTO 
6.5~\cite{lepto}, which includes order-$\alpha_S$ matrix elements
and models of higher-order QCD radiation.
The color-dipole model in ARIADNE 4.08~\cite{ariadne} 
provided a systematic check. 
The CTEQ4D parton distribution set~\cite{cteq4} with the Yang-Bodek correction,
Eq.~(\ref{eqn_by}), was used to evaluate the nominal CC cross section, and
the unmodified CTEQ4D PDF was used as an alternative PDF with 
smaller $d$-quark density.

Simulated resonant-state events were generated using 
PYTHIA 6.1~\cite{pythia}. States with masses between
150 and 280 GeV were simulated
in 10 GeV steps.  This program takes into account the finite width of
the resonant-state, but only includes the s-channel diagram. Initial-
and final-state QCD radiation from the quark and the effect of LQ 
hadronization before decay are taken into account, as is initial-state
QED radiation from the positron.

Generated events were input into a GEANT 3.13-based 
simulation~\cite{geant} of the
ZEUS detector. \ Trigger and offline processing requirements as used for the
data were applied to the simulated events.

\section{Event selection}
\label{sec_evsel} 

Events were selected with cuts similar to those used
in the CC cross-section measurement from the same data~\cite{ccpaper}.
The events were classified first according to $\gamma _0$, the hadronic 
scattering angle of the system relative to the nominal interaction point~\cite{ccpaper}.
If $\gamma _0$ was
sufficiently large, i.e. in the central region, tracks in the CTD were
used to reconstruct the event vertex. On the other hand, if $\gamma _0$
was small, i.e. in the forward region, the hadronic final state of such
$\nubar$-jet events was often outside the acceptance of the CTD, and thus
the vertex position was 
obtained from the arrival time of particles
entering the FCAL. The following selection cuts were then applied:

\begin{itemize}

\item to select high-mass $\nubar X$ states,
events were required to have substantial missing transverse momentum: 
$\ptmiss >20$~GeV;

\item a cut of $y<0.9$ discarded events in which the 
kinematic variables were poorly reconstructed;

\item events with $\ptmiss/E_{T}<0.4$ 
(where $E_T$ denotes the total transverse energy measured 
in the event) were removed to reject photoproduction 
background.  For events with $\gamma_0<0.4$, this cut was increased to
$0.6$;

\item NC background was removed by discarding events
with identified positrons;

\item non-$ep$ collision events caused by beam-gas, halo muons, 
and cosmic rays were removed by a series of standard cuts based on 
the general topology expected for events from $ep$ collisions originating 
from the interaction region
at the correct beam-crossing time.  

\end{itemize}

\begin{figure}[ptb]
\centerline{\epsfxsize=10cm  \epsfysize=11.4cm
\epsfbox{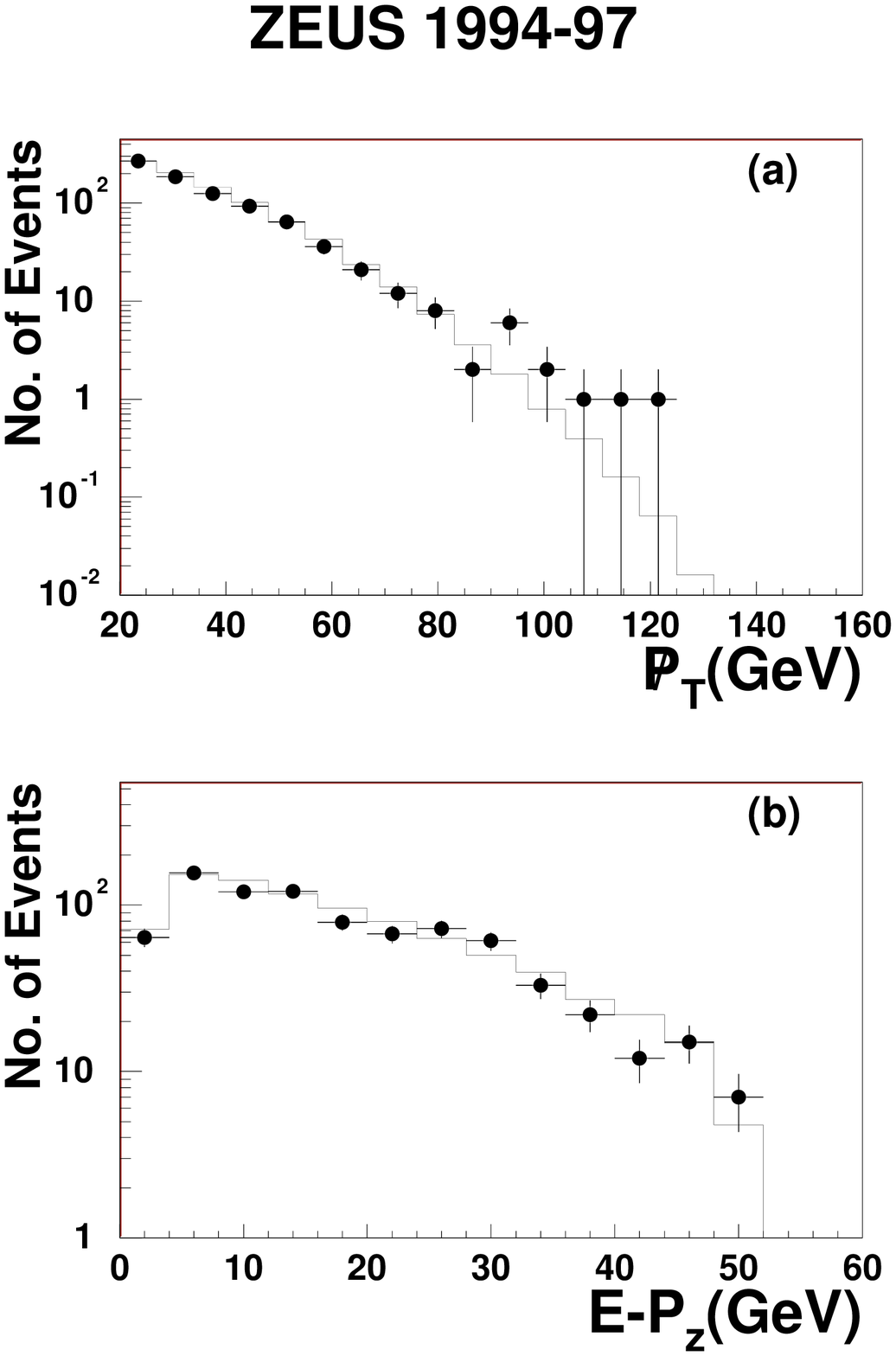}}\caption{
(a) The $\ptmiss$ distribution for
the final event sample. 
(b) The $(E-P_{Z})$ distribution.
The points represent the data and the histogram
is the SM MC prediction.
}
\label{plt_ptempz}
\end{figure}

The final sample
contains 829 events. \ 

The momentum carried by the antineutrino is extracted from the 
{$\ptmiss$} and the longitudinal momentum variable $(E-P_{Z})$ of the event;
distributions are shown in Fig.~\ref{plt_ptempz}.
The data and SM predictions agree except for 
{$\ptmiss > 90$~GeV}, 
where a slight excess is observed in the data.\ 
The $(E-P_{Z})$ distribution peaks near 10 GeV. These distributions are very
different from those of NC events, which
have small {$\ptmiss$} and an $(E-P_{Z})$
distribution peaked near twice the positron beam energy. These differences
arise from the undetected final-state antineutrino in this sample.

\begin{figure}[ptb]
\centerline{\epsfxsize=10cm  \epsfysize=11.4cm
\epsfbox{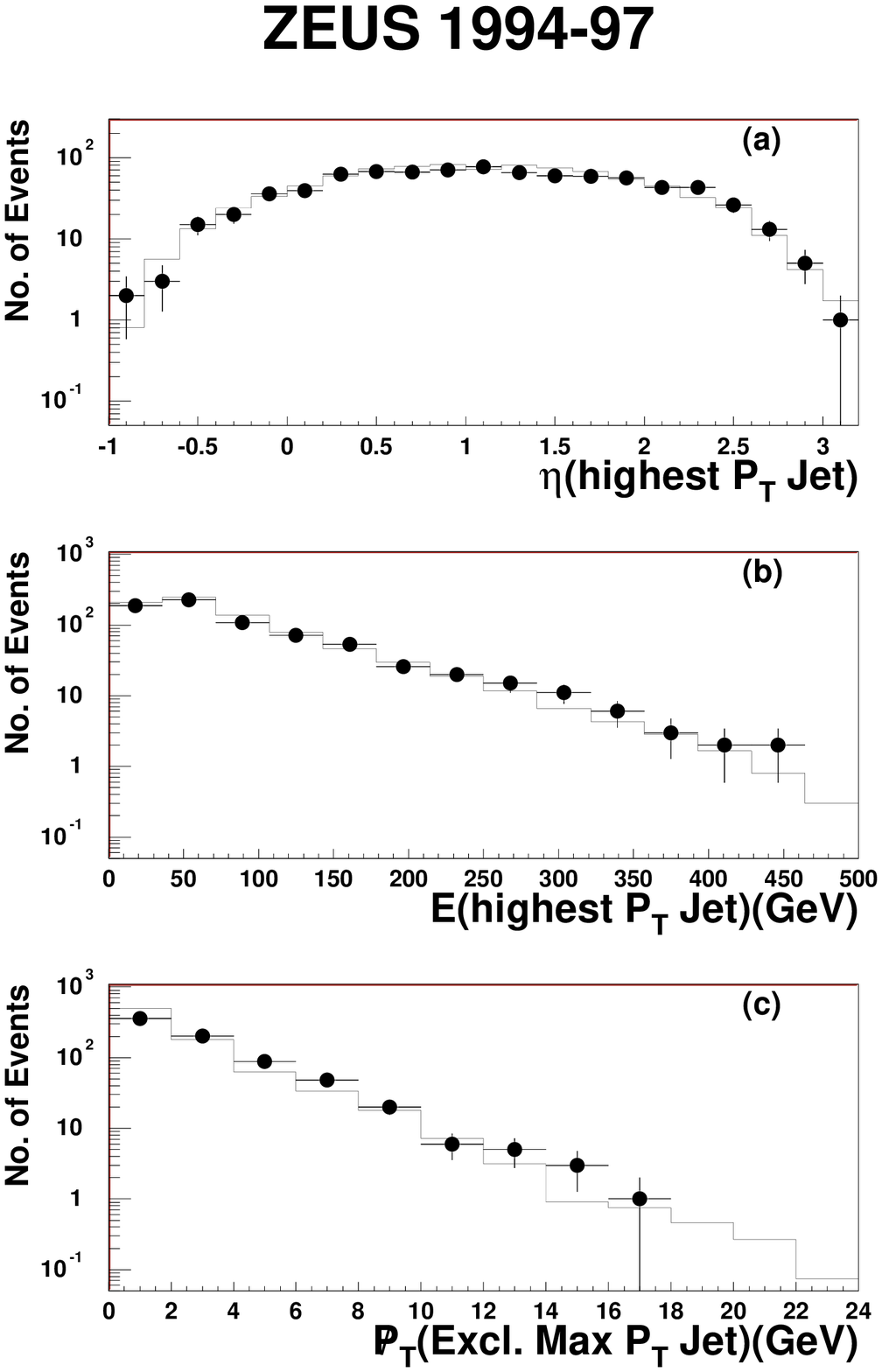}}\caption{Comparison of jet distributions in the data and
Monte Carlo.  (a) The pseudorapidity, $\eta$,
of the highest $P_T^j$ jet in each event.
(b) The energy of the highest $P_{T}^j$ jet.
(c) The missing transverse momentum
{$\ptmiss$} in each event when the highest
$P_{T}^j$ jet is excluded.
The points represent the data and the histogram the
SM Monte Carlo prediction.}
\label{plt_jets}
\end{figure}

Jets were identified
using the longitudinally-invariant $k_{T}$-clustering algorithm \cite{ktjets}
in inclusive mode \cite{ktjets2}.
At least one jet was required with transverse momentum 
$P_{T}^j>10\,\mbox{GeV}$. 
Fig.~\ref{plt_jets} shows the distributions of the pseudorapidity, $\eta$, of 
the highest $P_T^j$ jet~\footnote{The pseudorapidity is defined as 
$\eta = -\ln(\tan(\theta/2))$.}.
Also shown, for each event, are the energy of the highest $P_{T}^j$ jet 
and the $\ptmiss$ when the momentum of the highest $P_{T}^j$ jet is excluded. Reasonable agreement is observed between the data and SM predictions in each case.

The outer boundary of the
inner ring of FCAL towers was used to define a fiducial cut for the jet
reconstruction.  The centroid of the jet with the highest $P_T^j$ 
was required to be outside a $60\times 60$~cm$^2$ box on the face of the
FCAL centered on the beam pipe. This restricts the pseudorapidity 
of the jet to be less than roughly  $2.6$.  This
requirement removes 25 events,
bringing the total sample to 804 events.

\section{Mass and $\theta^\ast$ reconstruction}

\begin{figure}[ptb]
\centerline{\epsfxsize=10cm  \epsfysize=11.4cm
\epsfbox{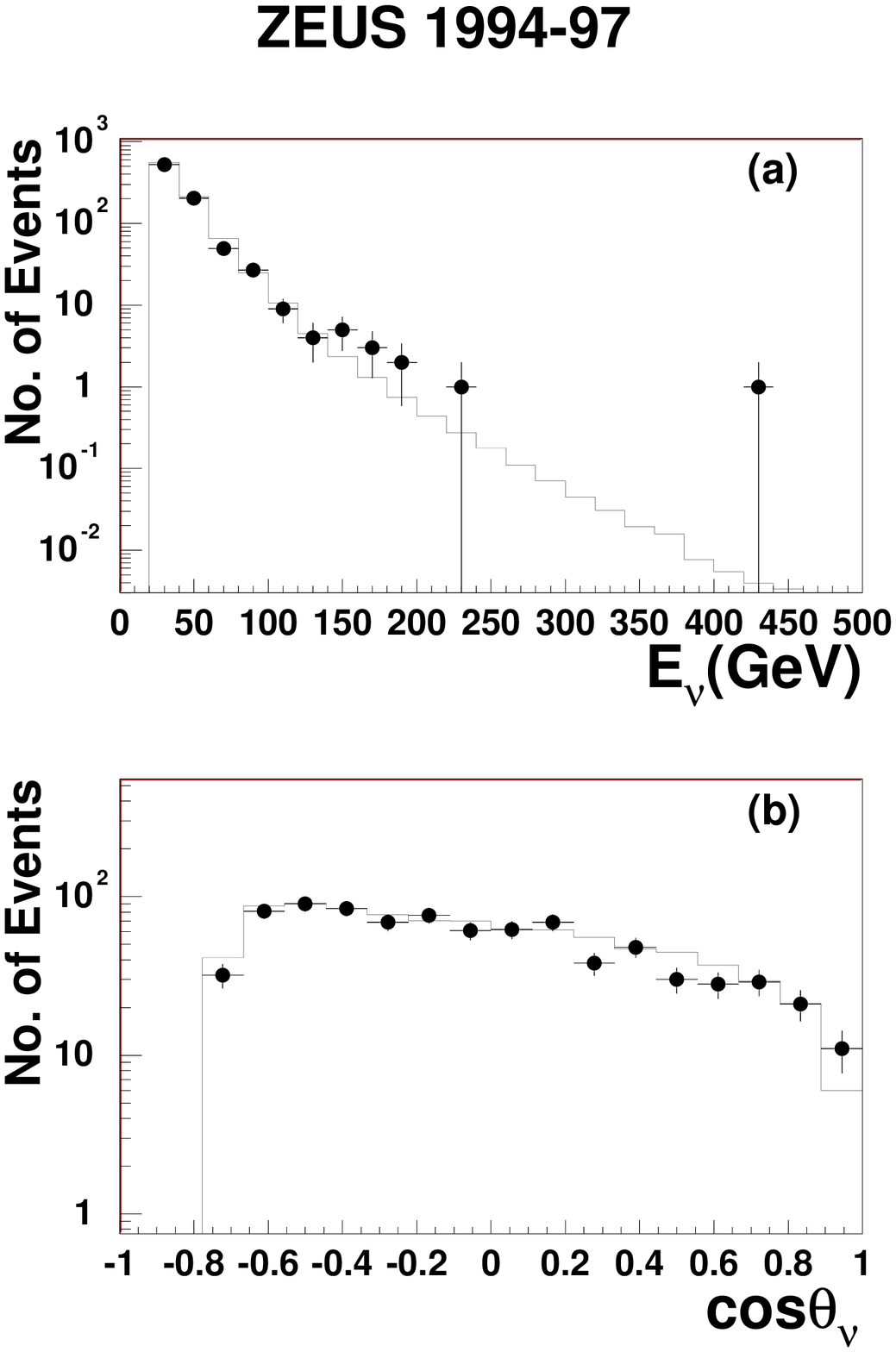}}\caption{
(a) The distribution of the energy of the final-state antineutrino in the 
lab frame, $E_{\nu}$.
(b) The distribution of $\cos \theta_{\nu}$, where $\theta_{\nu}$ is 
the polar angle of the scattered antineutrino in the lab frame. The 
forward direction ($\cos \theta_{\nu}=1$) 
corresponds to the proton beam direction.
The points are the data and the line is the SM Monte Carlo prediction. }
\label{fig_enutheta}
\end{figure}

It was assumed for the resonance search
that all the missing momentum is carried away by one 
antineutrino.
The invariant mass of the $\nubar$-jet system,~$M_{\nu j}$, 
was calculated using Eq.~(\ref{invm_eqn}) using only  the 
highest $P_{T}^j$ jet.
The jet direction was determined from the vector formed by the event
vertex and the jet centroid in the calorimeter.
The neutrino energy and angle were calculated as:
\begin{eqnarray*}
E_{\nu} & = & \frac{\ptmiss^2 + (E-P_Z)_{\nu}^2}{2\cdot(E-P_Z)_{\nu}}
\\
\cos \theta_{\nu} & = & 
\frac{\ptmiss^2 - (E-P_Z)_{\nu}^2}{\ptmiss^2 + (E-P_Z)_{\nu}^2}
\end{eqnarray*}
\noindent where $(E-P_Z)_{\nu}=2E_e-(E-P_Z)$.
Distributions of the reconstructed antineutrino energy and 
polar angle in the laboratory frame
($E_{\nu}$ and $\cos \theta_{\nu}$) are shown in Fig.~\ref{fig_enutheta}. 
Reasonable agreement is observed between data and the SM prediction.
Monte Carlo simulations of resonant states indicate that the
antineutrino energy and polar angle were measured with average resolutions of $16\%$ and $11\%$, respectively. The average systematic shift in $E_{\nu}$ was found to be less than $2\%$,  while the shift in $\theta_{\nu}$ was less than $1\%$.

Monte Carlo simulations of resonant states were used
to determine the resolution and estimate the possible bias for the
reconstructed mass. The mass resolution was obtained by performing a Gaussian
fit to the peak of the reconstructed mass spectrum. 
For resonant-state masses from 170 GeV to 270 GeV, the average mass resolution
was found to be $7\%$. The peak position of the Gaussian differed from the generated mass by less than $2\%$ over the entire range.

Note that energy-momentum conservation, 
assumed in order to calculate $E_{\nu}$ and $\theta_{\nu}$,
does not apply when undetected initial-state radiation (ISR) from the 
beam positron occurs. At high masses, QED 
radiation results in an underestimate of $E_{\nu}$ and 
an overestimate of $\theta_{\nu}$. This, as well as final-state QCD radiation, 
results in lower reconstructed masses,
leading to an asymmetry in the expected mass distribution. In a
simulation of a resonance of mass 220~GeV, only $1\%$ of events had an 
$M_{\nu j}$ 
more than $20\%$  higher than the true mass, while $16\%$ had an $M_{\nu j}$ 
more than $20\%$ lower than the true mass.

\begin{figure}[ptb]
\centerline{\epsfxsize=12cm  \epsfysize=14cm
\epsfbox{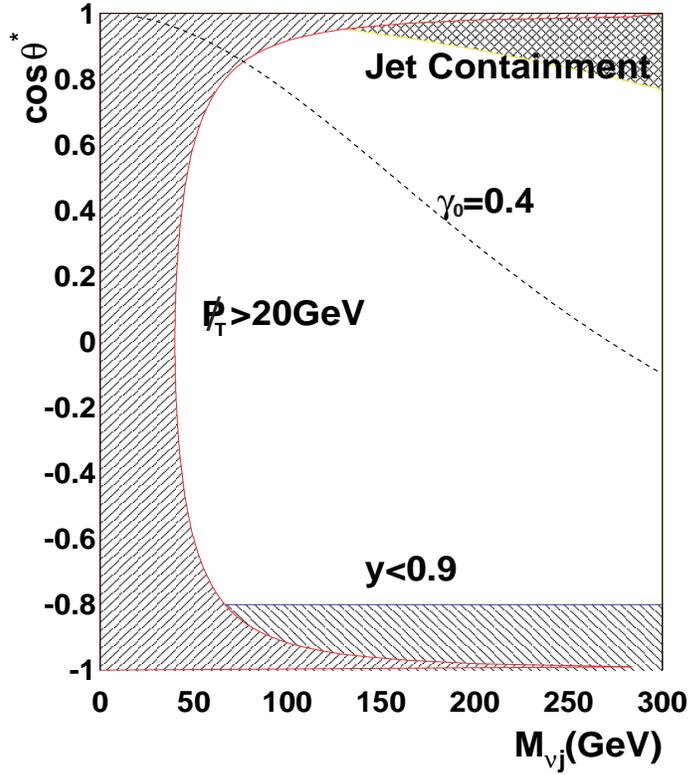}}\caption{Acceptance in the $\cos\theta^{\ast}$-$M_{\nu
j}$ plane. The shaded areas are the regions excluded by the requirements of
$\ptmiss>20$~GeV, $y<0.9$ and jet-containment
assuming an $eq \rightarrow \nubar q$ scattering at the nominal interaction 
point.   No detector simulation is included. 
The dotted $\gamma_0=0.4$ line shows the  
boundary between events using the FCAL
timing vertex (above) and the CTD tracking vertex (below).}
\label{plt_accep}
\end{figure}

In contrast to the resonance search, setting cross-section limits on 
 $e^{+}p\rightarrow\bar\nu X$ processes requires that a specific
production mechanism be assumed. For this reason, an invariant mass,
$M_{\nu j s}$, was calculated using all of the jets in the 
event with $P_{T}^j>10\,\mbox{GeV}$ and $\eta < 3$. Monte Carlo studies
show that, for narrow resonant states, using multiple jets gives more 
accurate mass reconstruction for events with more than one jet
(for masses above 150 GeV, $12\%$ of the simulated LQ events  have 
multiple jets).

The selection cuts described in Sect.~\ref{sec_evsel} determine the
kinematic region where mass reconstruction is possible. Figure~\ref{plt_accep}
shows the approximate regions in the 
$\cos\theta^{\ast}$-$M_{\nu j}$ plane which are
excluded by the requirements of $\ptmiss>20$~GeV, $y<0.9$ and the 
jet containment for events originating from
the nominal interaction point. In the unshaded regions, acceptance
is typically $\approx80\%$. The variable $\gamma$ denotes the scattering angle
of the struck quark. Events above the $\gamma_0=0.4$ line typically 
use the FCAL timing vertex, while those below this line use the vertex 
found from CTD tracking.

\section{Mass and $\cos \theta^\ast$ distributions}

\begin{figure}[ptb]
\centerline{\epsfxsize=12cm  \epsfysize=14cm
\epsfbox{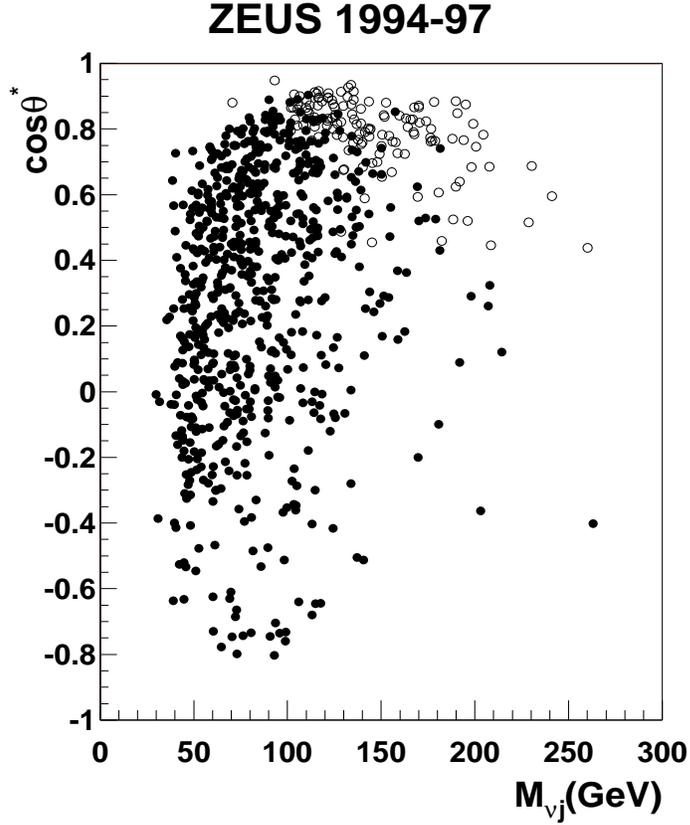}}\caption{The distribution of the final event sample in
the $M_{\nu j}$-$\cos\theta^{*}$ plane. Solid points indicate events 
reconstructed with a tracking vertex; open circles events reconstructed 
with a timing vertex. }
\label{plt_mscat}
\end{figure}

Figure~\ref{plt_mscat} shows the distribution of events in the $M_{\nu j}$
-$\cos\theta^{\ast}$ plane. 
The events populate the region of large acceptance described in
Fig.~\ref{plt_accep}.

\subsection{Systematic uncertainties}

The systematic uncertainties in the predicted rate of events
range from about $7\%$ at $M_{\nu j}\approx 
100\,\mbox{GeV}$ to about $20\%$ at $M_{\nu j}\approx 220\,\mbox{GeV}$, 
and over
$40\%$ at $M_{\nu j}\approx 260\,\mbox{GeV}$. \ The major sources of these
are uncertainties in the calorimeter energy scale, uncertainties in the 
simulation of the hadronic energy flow (established by
comparing results from the nominal
LEPTO\ MEPS\ model with a Monte Carlo sample using the alternative ARIADNE
model) and uncertainties in
the parton distribution functions.

Potential sources of systematic error which were found to
have negligible effects include 
reasonable variations of the selection cuts,
background-contamination uncertainties, timing-vertex uncertainties,
and the uncertainty in the luminosity determination. 

\subsection{Comparison with Standard Model}

\begin{figure}[ptb]
\centerline{\epsfxsize=10cm  \epsfysize=13cm
\epsfbox{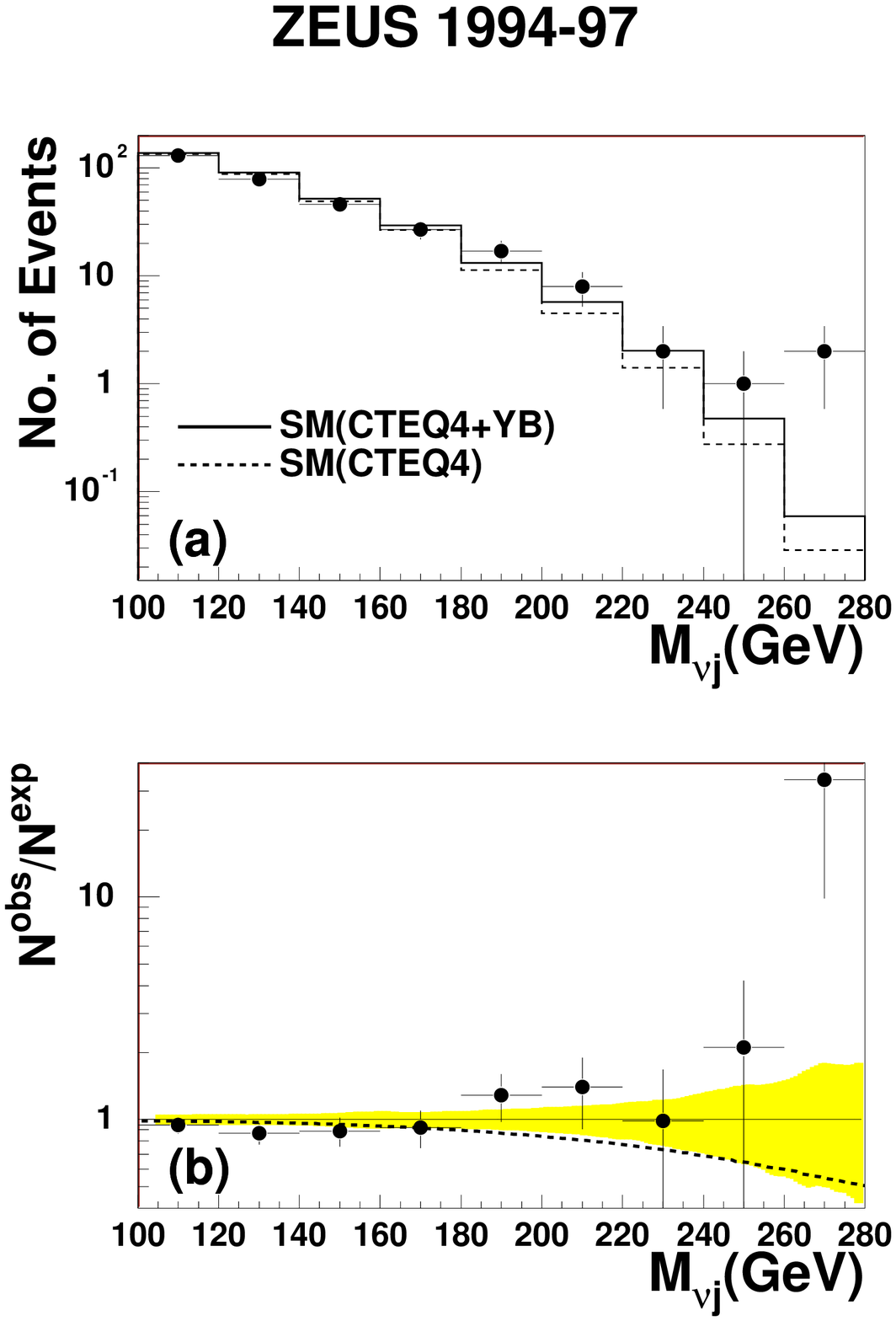}}
\caption{(a) The mass
distribution for the data (points) and Monte Carlo (histograms). The dashed
line shows the predicted mass spectrum when the CTEQ4D PDFs are used, while
the solid curve shows the distribution predicted when the $d$-quark density is
enhanced using the Yang-Bodek correction (see Eq.~(\ref{eqn_by})). 
(b) The ratio of the number of events observed to the number expected,
$N^{obs}/N^{exp}$, obtained using the Yang-Bodek correction. 
The shaded band indicates the systematic error in the
SM expectation. 
The dashed line shows the SM expectation when the Yang-Bodek correction
is not implemented. The error bars on the data points are calculated from
the square root of the number of events in the bin.
}
\label{plt_massres}
\end{figure}

\begin{figure}[ptb]
\centerline{\epsfxsize=10cm  \epsfysize=13cm
\epsfbox{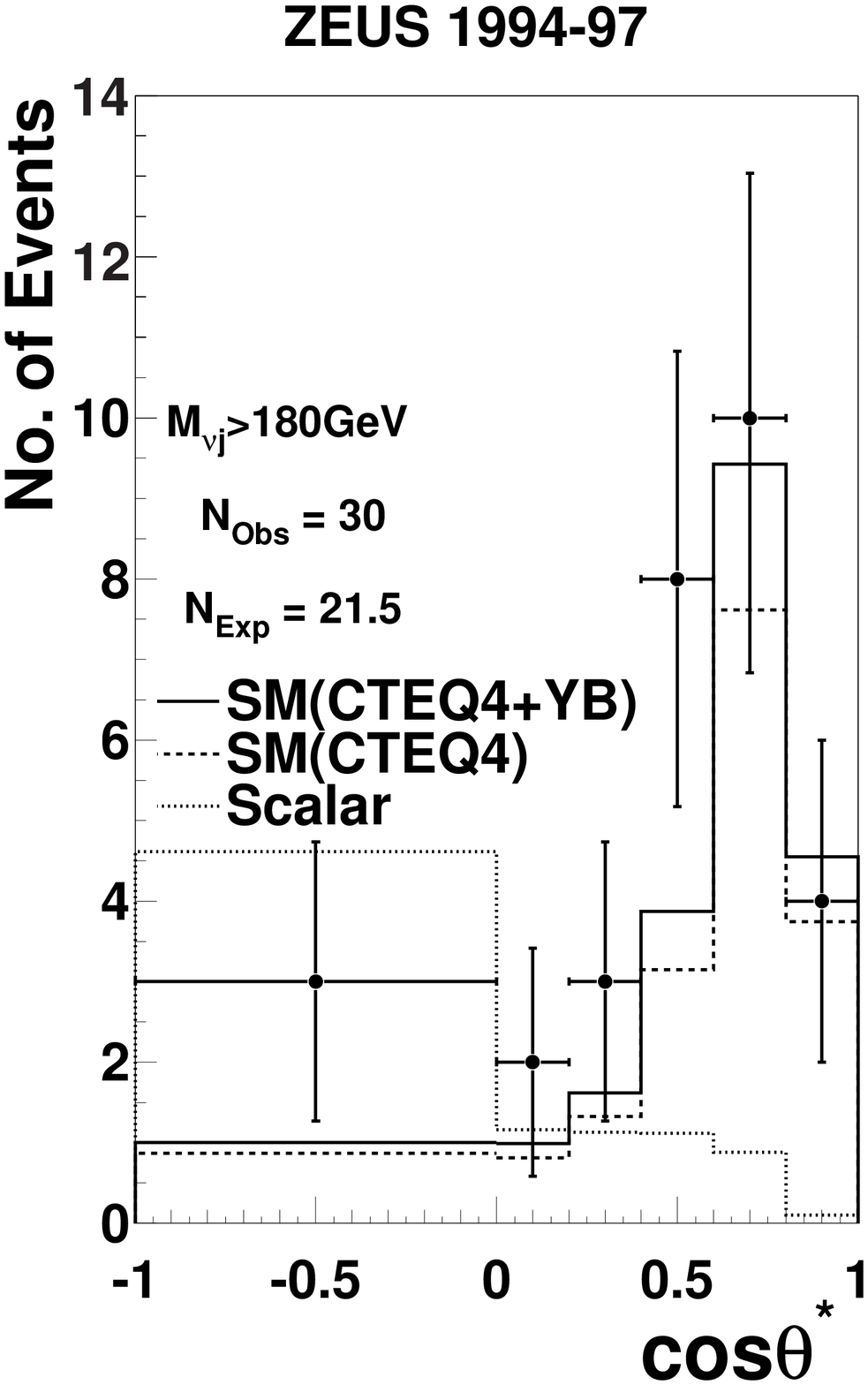}}
\caption{The $\cos\theta^{*}$ distribution of events with $M_{\nu j}>180 \; \mbox{GeV}$.  The dashed
line shows the predicted $\cos\theta^{\ast}$ spectrum when the CTEQ4D PDFs are used, while the solid curve shows the distribution predicted when the $d$-quark density is enhanced using the Yang-Bodek correction (Eq.~(\ref{eqn_by})).
Also shown is the
$\cos\theta^{\ast}$ distribution for a scalar resonance (dotted line) 
normalized to 9 events.
}
\label{plt_cth}
\end{figure}

In Fig.~\ref{plt_massres}(a), the observed mass distribution
is compared to the SM predictions from Monte
Carlo simulations using the CTEQ4D parton densities~\cite{cteq4} 
and the CTEQ4D PDF modified by the
Yang-Bodek correction of Eq.~(\ref{eqn_by}). 
The predictions using the CTEQ5~\cite{cteq5} or the NLO\ QCD
fit by Botje~\cite{botje} are similar to the modified CTEQ4D predictions. 
For $M_{\nu j}>180\;\mbox{GeV}$, the data tend to lie above the expectations.
There are 30 events observed in this region, while
$21.5\pm3.3$ are predicted ($16.0\pm2.4$ events for CTEQ4D without the
correction of Eq.~(\ref{eqn_by})). The uncertainty on the predicted
number of events is due to the effects described above.

Figure~\ref{plt_cth} shows the $\cos\theta^{\ast}$
distribution of the events with $M_{\nu j}>180 \; \mbox{GeV}$
together with the distribution expected for decay of a
narrow scalar resonance (normalized to 9 events). 
In the $\cos\theta^{\ast}<0.4$ region where the DIS background is 
suppressed, 8 data events are
observed while $3.6\pm 0.5$ SM events are expected. 

Given the limited statistics in the present data
and the systematic uncertainties of the SM predictions,
the observed mass spectrum is
compatible with SM expectations. 

\section{Limits on resonant-state production}
\label{sec_limits}

\begin{figure}[ptb]
\centerline{\epsfxsize=10cm  \epsfysize=13cm
\epsfbox{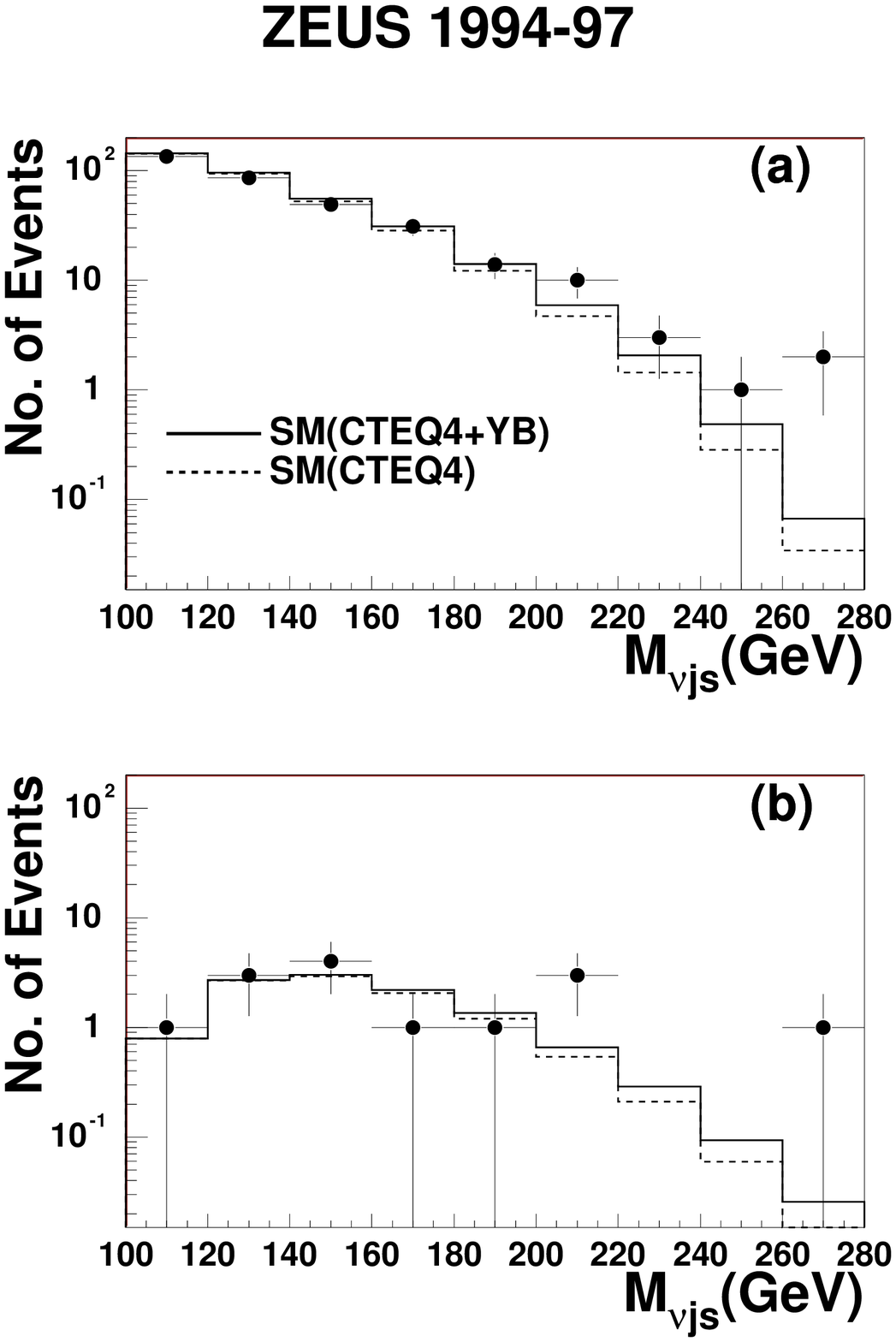}}
\caption{ a) The reconstructed mass spectrum using multiple jets 
for data (points) and SM expectation (histogram).
b) The mass spectrum using multiple jets after the cut ($\cos\theta^{\ast} < \cos\theta_{\max}^{\ast}$) for the scalar resonance search has been applied.
}
\label{plt_mjets}
\end{figure}

Since there is no evidence for a narrow resonance in the $\nubar$-jet 
data, limits may be set on the production of the resonant states 
listed in Table~\ref{tab_resst}.
Since such states would need to have a positron as well 
as an antineutrino decay channel, the cross-section limits
were set using these $\nubar$-jet data along with the 
$e^+$-jet data previously reported~\cite{ncref}.
Only couplings $\lambda \le 1$ are considered.
The limit-setting procedure assumes the states have the
same production and decay mechanism as the Monte Carlo used to generate
the resonance events. The invariant mass reconstructed using the
neutrino and all jets with $\ptmiss>10$~GeV and $\eta<3.0$,
$M_{\nu j s}$, was used to
set limits. 
The mass spectrum reconstructed with this technique
is shown in Fig.~\ref{plt_mjets}a, and is similar to that from
single jets (Fig.~\ref{plt_massres}).

The limit-setting procedure requires two parameters at each 
value of $M_{\nu j s}$: the mass window, $\Delta M_{\nu j s}$, 
and an upper cut ($\cos\theta_{\max}^{\ast}$) on
the measured value of $\cos\theta^{\ast}$. \ 
Simulations of both SM background and resonant signals
were used to find values for these parameters which
optimize observation of a signal relative to DIS background. \ 
For  a scalar resonance with a $\nubar$-jet final state, 
$\Delta M_{\nu j s}$ ranged from 20 to 35 GeV in the 160-280 GeV mass
range, while in the same range $\cos\theta_{\max}^{\ast}$ 
increased from 0.2 to 0.8. For a vector resonance
in the same $M_{\nu j s}$ range, $\Delta M_{\nu j s}$ increased from 15 to 35
GeV, while $\cos\theta_{\max}^{\ast}$ increased from 0.6 to 0.84. 
The mass spectrum after applying the optimal $\cos\theta^{\ast}$ cut for
the scalar search is shown in Fig.~\ref{plt_mjets}b.
A similar optimization procedure, performed for the $e^+$-jet final 
state using the NC data, has been described in a previous 
publication~\cite{ncref}.

To find the $95\%$ confidence level (CL)
upper limit on the resonant-state cross section, $\sigma_{\rm lim}$, 
a likelihood
is calculated using the Poisson probability for each decay channel:

\begin{equation}
  L_c(\sigma ) = e^{-(\mathcal{L} \beta_c \epsilon_c \sigma + 
N_c^{\rm bkg})}
 \frac{(\mathcal{L} \beta_c \epsilon_c \sigma + 
N_c^{\rm bkg})^{N_c^{\rm obs}}}{N_c^{\rm obs}!},
\label{eqn_lqlike}
\end{equation}

\noindent where $\mathcal{L}$ is the luminosity,
$\beta_c$ is the branching ratio of the decay channel, 
$N_c^{\rm obs}$ is the number of observed events, 
$N_c^{\rm bkg}$ is the expected number of DIS background, 
 and $\epsilon_c$ is the acceptance calculated from resonance Monte Carlo.
The subscript $c$ denotes the decay channel, which for this analysis
is either $\bar \nu q$ or $e^+q$, for the CC-like and NC-like final states, respectively. If more than one channel was used to set a limit, the likelihoods
for each channel were multiplied together to get the total likelihood, 
$L(\sigma )$. 
A flat prior probability density for the cross section $\sigma$ was assumed, 
such that the probability density, $f(\sigma)$, is
simply $f(\sigma) \propto L(\sigma)$.
A limit was then obtained on the cross section, 
$\sigma_{\rm lim}$, by solving:

\begin{equation}
\int_0^{\sigma _{\rm lim}}d\sigma f(\sigma ) = 
0.95 \int_0^{\infty}d\sigma f(\sigma )
\label{eqn_lqoptim}
\end{equation}

\noindent and the resulting cross-section limit was converted to a 
coupling limit $\lambda _{\rm lim}$ using the NWA (Eq.~(\ref{eqn_nwidth})).
Note that using two channels does not always produce a stronger limit
than using a single channel.

The limits on $\lambda$
depend on the accuracy of the NWA. 
Comparisons between the NWA and the full resonant-state cross sections 
show that the NWA was too high by up to a factor $1.7$ for 
$S_{e^+\bar u}$. This was corrected for in setting the limits.
For all other states, the NWA provides a reasonable approximation of the
full resonant-state cross section in the mass and coupling ranges studied.

\begin{figure}[ptb]
\centerline{\epsfxsize=12cm  \epsfysize=14cm
\epsfbox{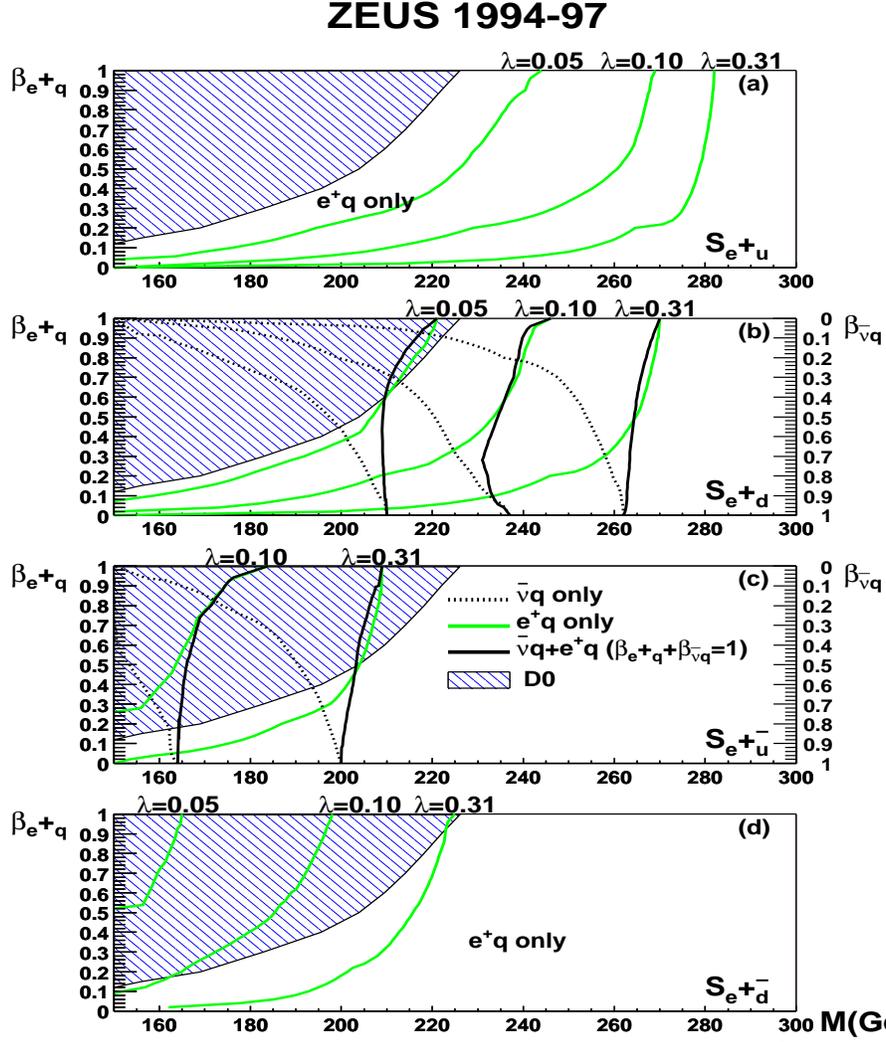}}\caption{
The branching ratios into $e^+q$ and $\nubar q$ 
(shown on the left and right axes, respectively)
vs. excluded mass for the scalar resonant 
states listed in Table~\ref{tab_resst}. 
For each limit curve, the area to the left of the curve is the excluded region.
Results for $e^+\ubar$,  $e^+\dbar$, $e^+u$ and $e^+d$ resonant states 
are shown for coupling strengths of $\lambda=0.05$,
$\lambda=0.10$, and $\lambda=0.31$.
The shaded region in each plot shows the mass range excluded by the D0 
experiment.
For (a)  $e^+u$ and (d) $e^+\dbar$ resonant states,
limits were set using only $e^+q$ data since $\nubar q$ decays are
forbidden by charge conservation.
The (b) $e^+d$ and (c) $e^+\ubar$ states 
have both $e^+q$ and $\nubar q$ decay channels.
The dotted line corresponds to
only $\nubar q$ data,
the shaded line corresponds to only $e^+q$ data, and the solid black line
corresponds to both the $e^+q$ and the $\nubar q$ data sets. 
The combined limits were calculated assuming that $\beta_{\nubar q}+
\beta_{e^+ q}=1$.
}
\label{fig_limscalar}
\end{figure}

\begin{figure}[ptb]
\centerline{\epsfxsize=12cm  \epsfysize=14cm
\epsfbox{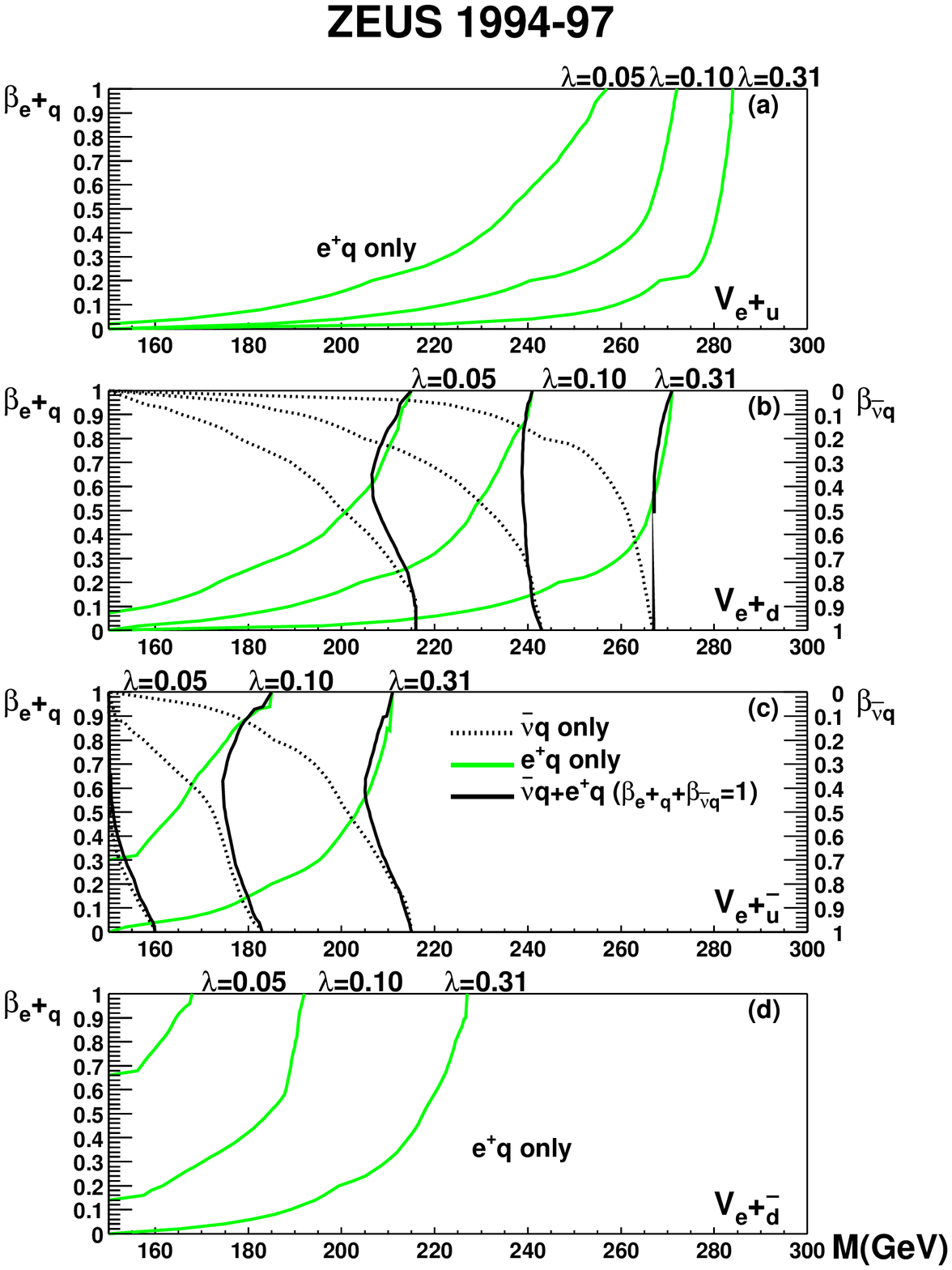}}\caption{
The branching ratios into $e^+q$ and $\nubar q$ 
(shown on the left and right axes, respectively)
vs. excluded mass for the vector resonant 
states listed in Table~\ref{tab_resst}. 
For each limit curve, the area to the left of the curve is the excluded region.
Results for $e^+\ubar$,  $e^+\dbar$, $e^+u$ and $e^+d$ resonant states 
are shown for coupling strengths of $\lambda=0.05$,
$\lambda=0.10$, and $\lambda=0.31$.
For (a)  $e^+u$ and (d) $e^+\dbar$ resonant states,
limits were set using only $e^+q$ data since $\nubar q$ decays are
forbidden by charge conservation.
The (b) $e^+d$ and (c) $e^+\ubar$ states 
have both $e^+q$ and $\nubar q$ decay channels.
The dotted line corresponds to
only $\nubar q$ data,
the shaded line corresponds to only $e^+q$ data, and the solid black line
corresponds to both the $e^+q$ and the $\nubar q$ data sets. 
The combined limits were calculated assuming that $\beta_{\nubar q}+
\beta_{e^+ q}=1$.
}
\label{fig_limvector}
\end{figure}

Figure~\ref{fig_limscalar} shows the limits obtained for the 
four scalar resonant states of Table~\ref{tab_resst} 
as a function 
of $\beta_{e^+ q}$ and $\beta_{\nubar q}$, 
the branching ratios into $e^+ q$ and $\nubar q$, respectively. 
The equivalent plots for vector resonant states are shown in 
Fig.~\ref{fig_limvector}.
The limits were 
calculated for coupling strengths of $\lambda=0.05$ and $\lambda=0.10$, 
as well as for coupling $\lambda=0.31\approx\sqrt{4\pi\alpha}$. 
For the $e^+u$ and $e^+\bar d$  resonances 
(a and d in Figs.~\ref{fig_limscalar} and~\ref{fig_limvector}), 
$\nubar q$ decays are forbidden
by charge conservation, so the limits are set using only the $e^+q$ channel.
The $e^+\bar u$ and $e^+d$ resonances (b and c) can provide both 
$e^+q$ and $\nubar q$ decays, so limits are calculated using
the $e^+$-jet and $\nubar$-jet data sets separately and combined. 
The combined $e^+ q \! + \! \nubar q$ limits, which assume 
$\beta_{\nubar q}+\beta_{e^+ q}=1$, 
are largely independent of branching ratio.
The limits obtained using only the  
$e^+$-jet (or the $\nubar$-jet) data 
allow for decay modes other than $e^+q$ and $\nubar q$, 
so the $e^+ q$ and the $\nubar q$ limits are applicable to a wider range 
of physics models 
than the combined $e^+q \! + \! \nubar q$ results.
The systematic uncertainties on the
predicted background described in Section~8.1 were found to change the 
excluded mass limits by less than $1\%$ for \mbox{$M_{\nu j}>220$ GeV}, and 
have therefore been neglected.  

The $e^+q$ and $\nubar q$ data have also been used to set limits on
scalar and vector resonances with second generation quarks. Assuming a
coupling strength of $\lambda = 0.31$ the mass limits for $e^+s$ states
decaying with 50~\% branching ratio to $e^+q$ and with 50~\% to 
$\nubar q$ are 207~GeV for a scalar and 211~GeV for a vector state.

For comparison, the limits on scalar
resonances obtained by the D0 experiment~\cite{D0_limits} 
at the Tevatron are shown by the shaded region. 
These limits are 
independent of both coupling and quark flavor.  Similar results to those
presented here have been published by the H1 experiment~\cite{H1results}.

\begin{figure}[ptb]
\centerline{\epsfxsize=10cm  \epsfysize=12cm
\epsfbox{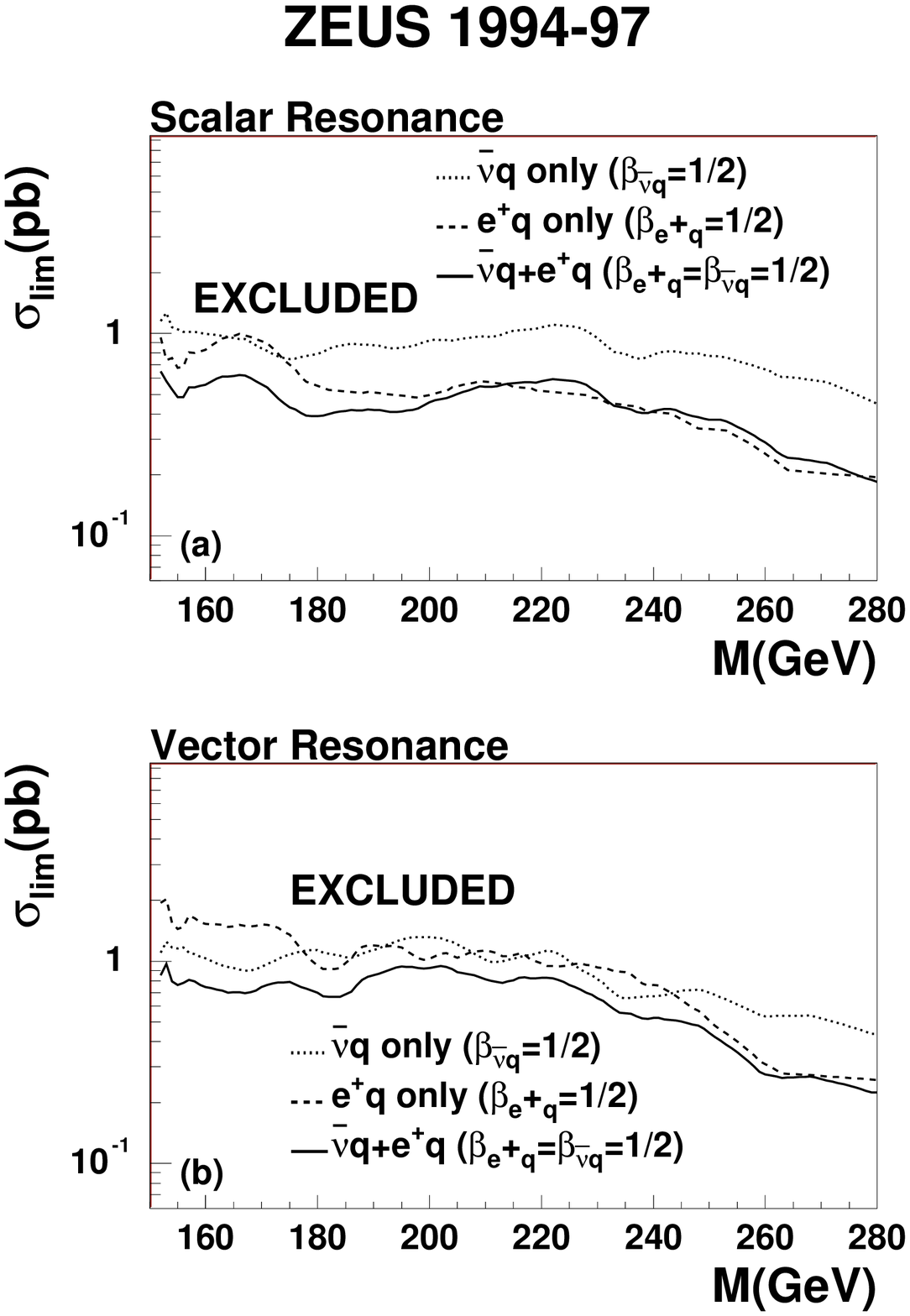}}\caption{
(a) Limits on the total production cross section
for a narrow scalar resonant state. 
(b) The corresponding limits for a narrow vector resonant state.
Limits derived from $e^+q$ ($\nubar q$) assume a branching ratio
$\beta_{e^+q}$ ($\beta_{\nubar q}$) of 1/2, while the
combined $e^+q \! + \! \nubar q$ limits assume branching
ratios of $\beta_{e^+q}=\beta_{\nubar q}=1/2$.
}
\label{plt_cseclimits}
\end{figure}

\begin{figure}[ptb]
\centerline{\epsfxsize=12cm  \epsfysize=14cm
\epsfbox{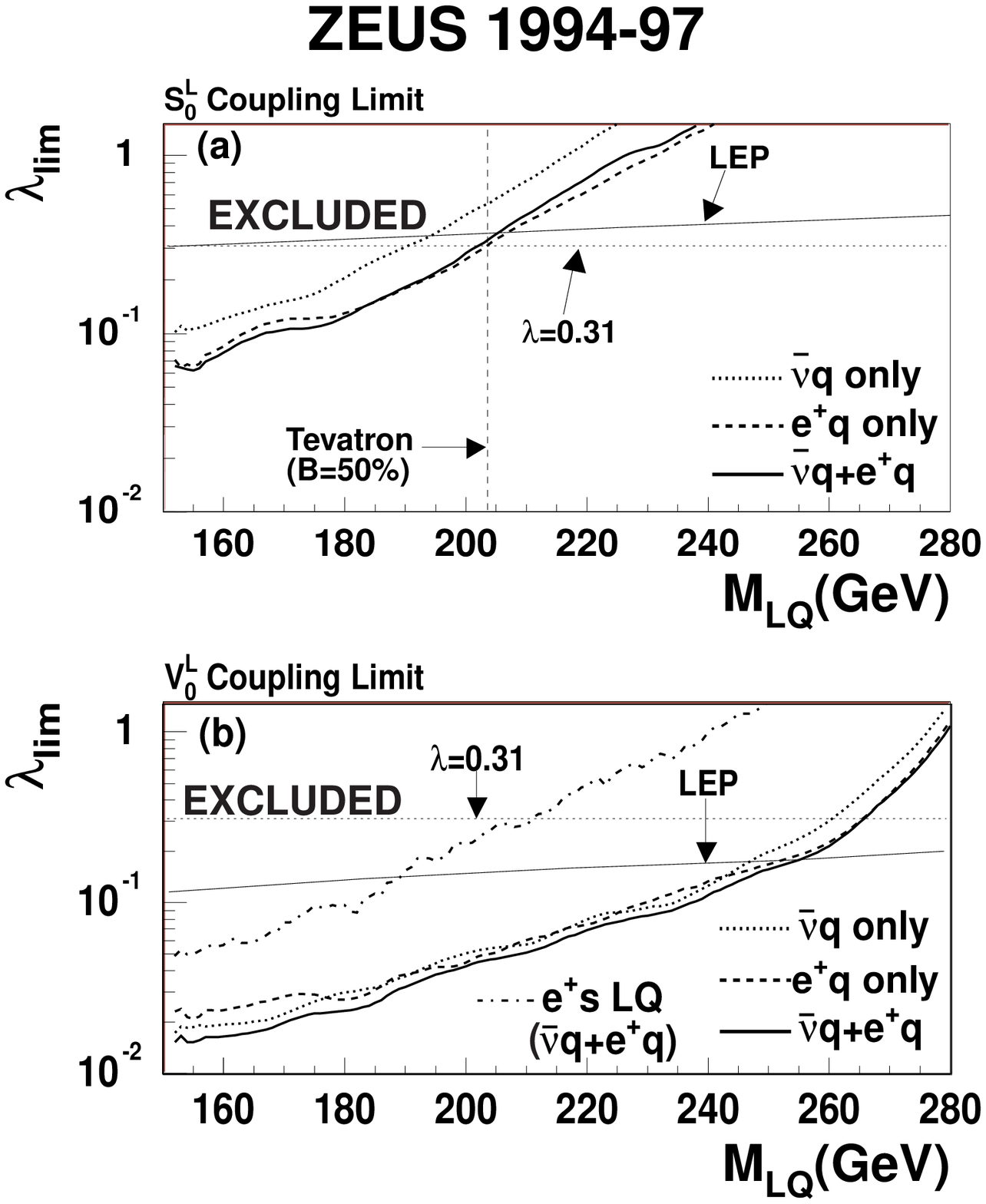}}
\caption[]{
(a) The limits on the coupling $\lambda_{\rm lim}$ for an $S_{0}^{L}$ LQ.
(b) The same for a $V_{0}^{L}$ LQ. 
Results from the $\nubar q$ and $e^+ q$ channels are shown, along with 
the limits obtained by combining the two channels. 
Also shown is the limit for second generation LQ's (dashed-dotted line).
In both plots, the horizontal line indicates the 
coupling $\lambda=0.31 \approx \sqrt{4 \pi \alpha}$.
For comparison, representative limits from the 
Tevatron~\cite{D0_limits} 
and LEP~\cite{OPAL_limits} are also shown.
}
\label{plt_lqlamlimits}
\end{figure}

\section{Model-dependent limits}

The limits on generic resonant states
were converted to limits on the production of LQ and squarks 
that have $e^+q$ and $\nubar q$ decays. 
Figure~\ref{plt_cseclimits}
shows the limit on the
production cross section, $\sigma_{\rm lim}$, 
for scalar and vector resonant states. 
Limits derived from $e^+q$ ($\nubar q$) assume a branching ratio
$\beta_{e^+q}$ ($\beta_{\nubar q}$) of 1/2, while the
combined $e^+q \! + \! \nubar q$ limits assume branching
ratios of $\beta_{e^+q}=\beta_{\nubar q}=1/2$.

\subsection{Leptoquarks limits}

The cross-section limits were converted to limits on leptoquark 
coupling using  Eq.~(\ref{eqn_nwidth}).
Figure~\ref{plt_lqlamlimits} shows the coupling limits for the 
$S_{0}^{L}$ and $V_{0}^{L}$ LQ species listed in Table~\ref{tab_lqs}.
If a coupling strength $\lambda=0.31 \approx\sqrt{4\pi\alpha}$
is assumed, the production of an $S_{0}^{L}$ LQ is excluded up to a mass of
204 GeV with $95$~\% CL, while the production of a $V_{0}^{L}$ LQ
is excluded up to a mass of 265 GeV. 
When the $\nubar q$ and $e^+q$ limits are combined, the resulting limits 
exclude approximately the
same mass range as the $e^+q$-only limit.
Also shown in Fig.~\ref{plt_lqlamlimits}
is the limit curve for second generation LQ's
of the type $V^0_L$ produced as an $e^+ s$ resonance.
The combined limits from $e^+ q$ and $\nubar q$ decays are shown. 
For comparison, limits from the D0 experiment with a  branching
ratio of $\beta_{e^+q}=1/2$ are shown~\cite{D0_limits}. Also included
are LQ limits from the OPAL experiment at LEP~\cite{OPAL_limits}.

\subsection{SUSY limits}

\begin{figure}[ptb]
\centerline{\epsfxsize=10cm  \epsfysize=12cm
\epsfbox{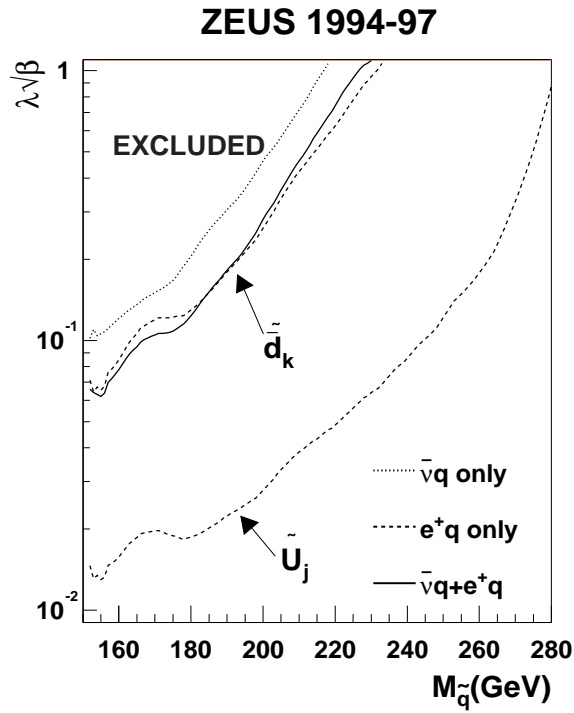}}\caption{
The limits on the coupling 
$\lambda \sqrt{\beta}$, where $\beta=\beta_{e^+q}+\beta_{\nubar q}$.
For the $\tilde{\bar{d}}_{k}$ squark, $\beta_{e^+ q}=\beta_{\nubar q}$, 
so results from the $\nubar$-jet  and $e^+$-jet channels are shown, 
along with the limit obtained by combining the two channels.
The $\tilde{u}_j$ limits are calculated using only $e^+q$ data since
the gauge invariance of the SUSY superpotential requires 
$\beta_{\nubar q}=0$.
}
\label{plt_limsusy}
\end{figure}

Limits were set on the production of the squarks 
listed in Table~\ref{tab_SUSY}. 
In addition to $\rpvio$
decays into $e^+q$ and $\nubar q$, squarks can also have 
$R_p$-conserving decays
into other final states. To remove the dependence on the 
branching ratios into these ${R}_p$-conserving states, 
limits were set on the quantity 
$\lambda \sqrt{\beta}$, where $\beta=\beta_{e^+q}+\beta_{\nubar q}$.
The limit-setting procedure does not account for possible contributions
to the $e^+$-jet and $\nubar$-jet channels from $R_p$-conserving decays.
Limits on 
$\tilde{{\bar{d}}}_{k}$ and $\tilde{u}_j$ are shown in Fig.~\ref{plt_limsusy}.
Because $\beta_{e^+ q}=\beta_{\nubar q}$ for the 
$\tilde{{\bar{d}}}_{k}$ decays, the combined $e^+ q \! + \! \nubar q$ 
limits are shown along with the limits obtained from the individual
decay channels.
For the  $\tilde{u}_j$ squark, 
$ \beta=\beta_{e^+ q}$ 
since  $\nubar q$ decays would violate gauge invariance.
Previous limits on $\rpvio$-squark production from smaller data sets
have been set by the H1 experiment~\cite{H1susy}.

\section{Conclusion}

A study of the $\bar\nu$-jet mass spectrum in
$e^{+}p\rightarrow\bar\nu X$ events at center-of-mass energy 300 GeV
has been performed with the ZEUS detector at HERA using
an integrated luminosity of $47.7\mbox{pb}^{-1}$.
Events with
topologies similar to high-$Q^{2}$ charged current DIS were selected. The
invariant mass, $M_{\nu j}$, was calculated from the jet with the highest 
transverse energy and
the antineutrino four-momenta. \ The jet momentum was measured directly, while
the antineutrino momentum was deduced from the energy-momentum imbalance
measured in the detector. No evidence for a narrow resonance was observed.
This analysis complements an earlier search for narrow resonances in
the $e^+$-jet final state.

In the absence of evidence for a high-mass resonant state, the $e^+$-jet 
and $\nubar$-jet
data sets were used to set limits on the production cross section of 
scalar and vector states decaying by either mode. 
Sensitivity to a resonant signal was optimized
by restricting the center-of-mass decay angle to remove most DIS
background and by choosing an appropriate mass window.
The resulting cross-section limits were converted to coupling limits 
on $e^+u$,~$e^+ d$,~$e^+ \bar u$ and $e^+ \bar d$ resonant states.

First-generation couplings between initial- and final-state 
quarks and leptons which conserve
flavor and electric charge were considered. Limits were calculated as a 
function of the $e^+q$ and $\nubar q$  branching ratios for small couplings 
and do not depend on a specific production mechanism.
For resonances with both $e^+q$ and $\nubar q$ decays, using both the
$e^+$-jet and $\nubar$-jet data gave limits which are largely independent
of the branching ratio if the state is assumed to have no additional decay modes.

The limits on generic resonant states were used to 
constrain the production of 
leptoquarks and $R_p$-violating squarks.
For leptoquark flavors whose branching ratios into $e^+q$ and $\nubar q$
are the same, exclusion limits of 204 GeV for
scalars and 265 GeV for vectors were obtained if 
a coupling strength $\lambda=0.31$ is assumed. 
Limits on the production of $\tilde{{u}}_j$ and $\tilde{{\bar{d}}}_k$ 
squarks were obtained directly from the limits on $e^+d$ and 
$e^+\bar u$ resonances, respectively. 

\section*{Acknowledgements}
We thank the DESY Directorate for their strong support and encouragement,
and the HERA machine group for their diligent efforts.
We are grateful for the  support of the DESY computing
and network services. The design, construction and installation
of the ZEUS detector have been made possible by the ingenuity and effort
of many people from DESY and home institutes who are not listed
as authors. It is also a pleasure to thank
W. Buchm\"uller, R. R\"uckl and M. Spira for useful discussions.

\end{document}

%% file: zeuspap_def.tex
\def\zero{{\scriptscriptstyle 0}}

\def\Z0{\ensuremath{Z^\zero}}

\def\SU2U1{{\rm SU}(2)\times{\rm U}(1)}

\def\max{{\rm max}}

\mathchardef\qsm=63
\mathchardef\pls=43
\mathchardef\mns=512
\mathchardef\plm=518
\mathchardef\eql=61
\mathchardef\smallleft=300
\mathchardef\smallright=301
\mathchardef\perslsh=47
\mathchardef\les=316
\mathchardef\gre=318
\mathchardef\leq=532
\mathchardef\grq=533
\chardef\usc=95
\chardef\til=126

\def\sqr#1#2#3{{\vcenter{\hrule height.#3ex\hbox{\vrule width.#2ex height#1ex
    \kern#1ex\vrule width.#3ex}\hrule height.#2ex}}}

\def\angleto{\vrule width.035em height2.1ex depth-.56ex\unskip\kern-.6ex\to}
\def\perchc#1{{\raise.4ex\hbox{$\mkern4mu#1{\it\perslsh}_
             {\mkern-5mu\scriptscriptstyle{{\rm o}\!{\rm o}}}^
             {\mkern-12.8mu\scriptscriptstyle{\rm o}}$}}}

\def\widebar#1{\mkern1.5mu\overline{\mkern-1.5mu#1\mkern-1.mu}\mkern1.mu}
\catcode`\@=11 
\def\parenbar{\mathpalette\p@renb@r}
\def\p@renb@r#1#2{\vbox{%
  \ifx#1\scriptscriptstyle \dimen@.7em\dimen@ii.2em\else
  \ifx#1\scriptstyle \dimen@.8em\dimen@ii.25em\else
  \dimen@1em\dimen@ii.4em\fi\fi \offinterlineskip
  \ialign{\hfill##\hfill\cr
    \vbox{\hrule width\dimen@ii}\cr
    \noalign{\vskip-.3ex}%
    \hbox to\dimen@{$\mathchar300\hfil\mathchar301$}\cr
    \noalign{\vskip-.3ex}%
    $#1#2$\cr}}}
\catcode`\@=12 

\def\dbar{\ensuremath{\widebar{d}}}
\def\ubar{\ensuremath{\widebar{u}}}

\def\nubar{\ensuremath{\widebar{\nu}}}

\newbox\struttbox
\setbox\struttbox=\hbox{\vrule height1.65ex depth.485ex width0pt}
\def\strutt{\relax\ifmmode\copy\struttbox\else\unhcopy\struttbox\fi}
\def\stru#1#2{\relax\ifmmode\hbox{\vrule height#1 depth#2 width0pt}
\else\vrule height#1 depth#2 width0pt\fi}

\def\ronum#1{\uppercase\expandafter{\romannumeral#1}}
\def\ronuml#1{\expandafter{\romannumeral#1}}

\DeclareMathAlphabet{\mathbf}{OT1}{cmr}{bx}{sl}

%% file: zeuspap_fmt.tex
 {\end{list}}
 {\end{list}}
 {\end{list}}

\catcode`\@=11 
\newlength{\@fninsert}
\newlength{\@fnwidth}
\renewcommand{\@makefntext}[1]%
  {\noindent\makebox[\@fninsert][r]{\@makefnmark}\hfil%
  \parbox[t]{\@fnwidth}{#1}}
\catcode`\@=12 
\newlength{\localtextwidth}
\catcode`\@=11 
\newsavebox{\tmpbox}
\newlength{\@captionmargin}
\newlength{\@captionwidth}
\newlength{\@captionitemtextsep}
\catcode`\@=11 
\renewcommand\section{\@startsection{section}{1}{\z@}%
                                   {-3.5ex \@plus -1ex \@minus -.2ex}%
                                   {2.3ex \@plus.2ex}%
                                   {\normalfont\Large\bfseries}}
\renewcommand\subsection{\@startsection{subsection}{2}{\z@}%
                                   {-3.25ex\@plus -1ex \@minus -.2ex}%
                                   {1.5ex \@plus .2ex}%
                                   {\normalfont\large\bfseries}}
\renewcommand\subsubsection{\@startsection{subsubsection}{3}{\z@}%
                                   {-3.25ex\@plus -1ex \@minus -.2ex}%
                                   {1.5ex \@plus .2ex}%
                                   {\normalfont\large\bfseries}}
\renewcommand\paragraph{\@startsection{paragraph}{4}{\z@}%
                                   {3.25ex \@plus1ex \@minus.2ex}%
                                   {1.2ex \@plus .2ex}%
                                   {\normalfont\normalsize\bfseries}}
\catcode`\@=12 

\newsavebox{\sesbox}
\newlength{\seslen}